\documentclass[10pt,journal,compsoc]{IEEEtran}
\ifCLASSOPTIONcompsoc
  \usepackage[nocompress]{cite}
\else
  \usepackage{cite}
\fi
\ifCLASSINFOpdf
\else
\fi
\usepackage{amsmath, amssymb, cite, algorithmic, array, url, gensymb,enumitem,bm, multirow,multicol, graphicx}
\newcommand{\mli}[1]{\mathit{#1}}
\renewcommand{\d}[1]{\ensuremath{\operatorname{d}\!{#1}}}

\DeclareMathOperator*{\argmin}{arg\,min}

\usepackage{array}

\ifCLASSOPTIONcompsoc
  \usepackage[caption=false,font=footnotesize,labelfont=sf,textfont=sf]{subfig}
\else
  \usepackage[caption=false,font=footnotesize]{subfig}
\fi
\begin{document}
%
\title{Two-pass Light Field Image Compression for Spatial Quality and Angular Consistency}
%
%
%
%

\author{Bichuan~Guo,~\IEEEmembership{Member,~IEEE,}
        Jiangtao~Wen,~\IEEEmembership{Fellow,~IEEE,}
        and~Yuxing~Han,~\IEEEmembership{Member,~IEEE}
\IEEEcompsocitemizethanks{\IEEEcompsocthanksitem B. Guo and J. Wen are with the Department
of Computer Science and Technology, Tsinghua University, Beijing,
China, 100084.
\IEEEcompsocthanksitem Y. Han is with the Department of Automation, South China Agricultural University, Guangzhou, China, 510642.
\IEEEcompsocthanksitem This work was supported by the National Natural Science Foundation of China (Project Number 61521002).}
}

\IEEEtitleabstractindextext{%
\begin{abstract}
The quality assessment of light field images presents new challenges to conventional compression methods, 
as the spatial quality is affected by the optical distortion of capturing devices,
and the angular consistency affects the performance of dynamic rendering applications.
In this paper, we propose a two-pass encoding system for pseudo-temporal sequence based
light field image compression with a novel frame level bit allocation framework that 
optimizes spatial quality and angular consistency simultaneously.
Frame level rate-distortion models are estimated during the first pass,
and the second pass performs the actual encoding with optimized bit allocations
given by a two-step convex programming.
The proposed framework supports various encoder configurations.
Experimental results show that comparing to the anchor HM 16.16 (HEVC reference software), 
the proposed two-pass encoding system on average
achieves 11.2\% to 11.9\% BD-rate reductions for the all-intra configuration, 
15.8\% to 32.7\% BD-rate reductions for the random-access configuration,
and 12.1\% to 15.7\% BD-rate reductions for the low-delay configuration.
The resulting bit errors are limited,
and the total time cost is less than twice of the one-pass anchor.
Comparing with our earlier low-delay configuration based method,
the proposed system improves BD-rate reduction by 3.1\% to 8.3\%,
reduces the bit errors by more than 60\%, and achieves more than 12x speed up.
\end{abstract}

}

\maketitle

\IEEEdisplaynontitleabstractindextext

%
\IEEEpeerreviewmaketitle

\IEEEraisesectionheading{\section{Introduction}\label{sec:introduction}}
%
%
%
%
\IEEEPARstart{T}{he} concept of light field was introduced by A. Gershun \cite{Gershun}
back in 1939. In contrast to traditional 2-D images, 
light field describes the intensity of light rays 
that pass through each point $(V_x, V_y, V_z)$ in space, at each possible direction $(\theta,\phi)$, 
wavelength $\lambda$ and time $t$, enjoying a total of 7 degrees of freedom.
With the assumption that the light intensity remains constant on any light trajectory, 
a 4-D light field image (LFI) can be used to capture a snapshot of the light field at any moment.
The 4-D snapshot can be thought of as light rays coming from 2-D points on the focal $(xy)$ plane to 2-D points on the aperture $(uv)$ plane.
\begin{align}
\mathcal{L} = \mathcal{L}(u,v,x,y).	
\end{align}

Traditionally, LFIs are captured by multi-camera arrays, 
each camera capturing a particular perspective, known as a sub-aperature image (SAI) of the scene.
Recent developments in capturing technology allowed LFI capturing by micro-lense arrays with conventional image sensors,
 where each micro-lense acts as an individual low resolution camera.
This technique allowed the manufacturing of commercially available light field cameras \cite{cfp-icme},
greatly expanding the application of LFIs.
The LFI contains complete information of the light rays in a scene, therefore, 
it can be used to render new views with changed viewpoint and changed focal point.
A framework for creating virtual renders by resampling and interpolation of captured pixels was proposed by M. Levoy \cite{levoy1996light}.

M. Rerabek \textit{et al.} \cite{rerabek2016new} proposed an LFI dataset captured by a Lytro Illum \cite{lytro} camera, 
the resulting resolution of the 4-D LFIs is 15$\times$15 for the $uv$ plane, and 434$\times$625 for the $xy$ plane, producing 61 million pixels for a single LFI.
Due to the high information nature of LFI, 
efficient algorithms are required to compress the information of the sheer number of pixels created during acquisition.
The raw data from camera sensors are converted into a device-specific format after pre-processing including de-mosaicing and de-vignetting \cite{viola2017comparison}.
Two main approaches for LFI compression have been proposed in literature using modern image/video coding standards for conventional, non-LF image and video.
The first approach makes use of self-similarity compensated prediction,
usually implemented as an add-on to video compression standards,
to exploit the inherent spatial correlation in LFIs.
The second approach, referred to as pseudo-temporal sequence (PTS) based compression,
decomposes the LFI into multiple SAIs and arranges them into a sequence,
which is then compressed with conventional video encoders (e.g. H.265/HEVC)
 with motion estimation/compensation.
Many arrangement schemes and coding structures have been proposed.
I. Viola \textit{et al.} \cite{viola2017comparison} compared multiple compression algorithms based on these two approaches,
and showed that PTS-based compression consistently performs better.

Most LFI compression research is based on objective evaluations using the peak signal-to-noise ratio (PSNR) metric.
Original pixels and reconstructed pixels from decompression are compared,
equal weights are assigned to all coordinates $(u,v,x,y)$.
However, LFIs captured by micro-lense arrays often have severe optical distortion at various regions of the 4-D space.
The physical structure of micro-lense arrays often causes pixels in most external SAIs to be dimmed or even completely black.
This phenomenon is observed in most images from \cite{rerabek2016new}.
In \cite{viola2017comparison}, I. Viola \textit{et al.} dealt with such distortion by removing the most external SAIs.
To quantitatively measure this distortion, a fourth channel,
in addition to conventional RGB channels, is given in the dataset \cite{rerabek2016new}.
This channel represents the confidence that the captured pixel is a faithful sample of the light ray intensity.
Coding distortion of high confidence pixels incurs larger loss comparing to low confidence pixels,
since undistorted pixels carry more reliable information.
Therefore,
different weights shall be given to pixels with different confidence to account for the effect of optical distortion. 

Another issue in compression quality assessment is related to virtual renders.
To create re-focused images, the rendering procedure involves interpolation of adjacent pixels \cite{levoy1996light}.
If a pixel has a large distortion from compression, the interpolation result will be affected even if its adjacent pixels are accurately reproduced.
Moreover, if the coding distortion among adjacent pixels fluctuates wildly,
the quality of rendered virtual views will be sensitive to small adjustments of the chosen viewpoint and focal point.
These two facts suggest that adjacent pixels should have similar coding distortion
to improve the overall rendering quality.
For PTS-based compression, this requirement is satisfied when pixels with identical $(u,v)$ coordinates are adjacent in $(x,y)$,
as they are almost always in the same coding unit of the same PTS frame.
However, interpolated pixels with identical $(x,y)$ coordinates and are adjacent in $(u,v)$ belong to different PTS frames,
and can have a large difference in coding distortion.
G. Wu \textit{et al.} \cite{wu2017light} coined the term angular consistency (since the $uv$ space is also called the angular space) to describe this phenomenon,  which can not be measured by the PSNR.

To tackle these issues, this paper proposes an optimization target 
where the spatial quality affected by optical distortion and the angular consistency are incorporated into a unified function.
A frame level bit allocation framework for PTS-based LFI compression
is then proposed,
which allocates the total bits to SAIs efficiently
according to the proposed optimization target.
The main contributions of this work are:
\begin{enumerate}
	\item  A weighted distortion model that assigns weights to coding distortion according to pixel confidence;
	\item A smoothness penalty term that measures the angular consistency in terms of coding distortion differences among adjacent pixels in the angular space;
	\item A generic optimization target based on maximum a posteriori (MAP) criteria that combines spatial quality and angular consistency;
	\item Frame level rate-distortion models that relate frame bit costs with coding distortion for PTS-based LFI compression, adapted to various encoder configurations;
	\item A two-step strategy that solves the optimization problem with convex programming, which yields efficient frame level bit allocations;
	\item A two-pass encoding system that estimates the parameters of the proposed rate-distortion models, and conducts encoding according to the deduced bit allocation.
\end{enumerate}

Earlier results of the on-going research has already been published in \cite{guo2018convex}.
Compared with the earlier results, in this paper, the following extensions are made:
\begin{enumerate}
	\item A theoretical analysis and justification of the proposed optimization target;
	\item Rate-distortion models that incorporate inter-frame dependency, which can be adapted to various encoder configurations (all-intra/random-access/low-delay);
	\item A two-pass parallel compression scheme that speeds up the original multi-pass serial trial compressions.
\end{enumerate}
Experimental results show that compared with our previous work,
the proposed two-pass encoding system achieves 3.1\% to 8.3\% further BD-rate improvement, 
reduces the bit errors by more than 60\%, 
and speeds up the encoding processes by more than 12x.

The rest of the paper is organized as follows.
An overview of the related work is presented in Section \ref{sec:related-work}.
Section \ref{sec:opt-target} will provide a detailed description of the weighted distortion model, 
the smoothness penalty and the optimization target.
Section \ref{sec:alloc} will introduce the frame level bit allocation framework for PTS-based LFI compression. 
The frame level rate-distortion models 
and the deduced optimization problem will be explained in detail.
The two-pass LFI compression scheme 
will also be proposed.
Experimental results of the proposed algorithms are given in Section \ref{sec:exp},
and Section \ref{sec:conclusion} concludes the paper.

\section{Related Work} \label{sec:related-work}
\subsection{LFI compression} \label{subsec:related-work-1}
Most literature categorizes LFI compression methods into two categories,
as mentioned in Section \ref{sec:introduction}.
The first approach, based on self-similarity, relies on the fact that
various parts of the LFI are highly correlated.
The concept of self-similarity was first introduced by Conti \textit{et al.} \cite{conti2011spatial},
where self-similarity compensated prediction was integrated into the H.264/AVC  \cite{wiegand2003overview} standard
to efficiently compress LFIs.
Self-similarity compensated prediction shares similar characteristics with intra block copy (IBC) in H.265/HEVC \cite{sullivan2012overview} screen content coding.
Later this method is incorporated into H.265/HEVC to further improve coding efficiency \cite{conti2016hevc}.
More improvements, including bi-directional self-similarity compensated prediction \cite{conti2016hevc2}
and template matching \cite{monteiro2016light}, have been proposed.
Y. Li \textit{et al.} \cite{li2016coding} proposed the displacement intra prediction, 
which introduced inter-prediction scheme into H.265/HEVC intra prediction to further exploit self-similarity.

The second approach is based on PTS.
A light field image is first decomposed into multiple SAIs, 
which are arranged into a sequence according to a chosen order.
Then, modern video coding standards can be utilized to
remove the correlation among SAIs.
Many arrangement schemes have been proposed, including spiral scan order \cite{viola2017comparison}, 
raster scan order \cite{perra2016high}, and a circular sequencing approach\cite{hariharan2017low}.
However, most schemes were proposed in an ad hoc manner, 
while the bit allocation among SAIs and rate-distortion performance were loosely managed.
D. Liu \textit{et al.} \cite{liu2016pseudo} proposed a 2-D hierarchical coding structure 
that imitates the physical structure of micro-lense arrays.
The decomposed SAIs were grouped into four quadrants, 
where each SAI can be referenced by adjacent SAIs both horizontally and vertically.
This significantly improved the coding efficiency.
L. Li \textit{et al.} \cite{li2017pseudo} further proposed a bit allocation scheme by adapting the R-$\lambda$ model in H.265/HEVC to the 2-D hierarchical coding structure.

\subsection{LFI quality assessment} \label{subsec:related-work-2}
G. Wu \textit{et al.} \cite{wu2017light} summarized LFI quality assessment methods in terms of 
spatial-angular resolution, spatial quality and angular consistency.
Q. Fu \textit{et al.} \cite{fu2011image} compared light field cameras to conventional cameras,
showing that due to the re-focusing rendering technique of LFI,
light field cameras exhibit more stabilized visual resolution in terms of
depth of field.
Conventional objective metrics such as PSNR and structure similarity index (SSIM) \cite{wang2004image} have been used in the literature to assess the spatial quality of LFIs.
They were applied both to the SAIs and to the synthesized virtual views. 
V. Adhikarla \textit{et al.} \cite{adhikarla2017towards} presented an interactive light field viewing setup for the subjective evaluation of angular consistency.
They also evaluated the performance of many conventional objective metrics, including PSNR, SSIM, VQM \cite{wolf2007application}  etc., on LFIs.

\subsection{Frame level rate control} \label{subsec:related-work-3}
Most video rate control algorithms are based on the assumption that instantaneous bit-rate should remain relatively constant to ensure a stable video playback quality.
However, this assumption is no longer valid for PTS-based LFI compression.
Although the decomposed SAIs are arranged into a sequence,
the encoded video bit-stream is never played in the temporal order.
Instead, the 4-D LFI is reconstructed from the entire bit-stream to enable further processing such as re-focused virtual rendering.
This feature is similar to the rate control of variable bit-rate (VBR) coding,
where the instantaneous bit-rate is allowed to have more fluctuations compared with constant bit-rate (CBR) coding.
In order to gain knowledge about the optimal bit allocation in VBR coding,
many VBR algorithms \cite{yin2004practical}\cite{que2006efficient} employ a two-pass procedure.
In the first-pass, also called the trial compression, 
encoding related statistics, 
especially the rate-distortion behavior of each frame, is collected.
The second-pass uses the collected statistics to perform the actual encoding, with adjusted bit allocation.
However, the rate-distortion behavior of a frame is related to previously encoded frames,
resulting in an exponential complexity in modeling.
C.-Y. Hsu \textit{et al.} \cite{hsu1997joint} proposed an iterative framework,
where the rate-distortion behavior of each frame is assumed to be mutually independent,
and the encoding process is run iteratively to seek convergence.
C. Pang \textit{et al.} \cite{pang2011frame} proposed a frame level rate-distortion model that features backward reference, 
where a finite number of past frames are included in the rate-distortion model of the current frame.
In our previous work \cite{guo2018convex}, we used the iterative framework from \cite{hsu1997joint}.
In this paper we propose a novel GOP-based rate-distortion model to address the issue of inter-frame dependency.
This allows us to achieve faster encoding while maintaining the coding efficiency.

Once the frame level rate-distortion behavior is known,
the task of allocating bits efficiently to yield the best overall compression quality can be formulated as an optimization problem 	with a given constraint on total bit cost.
This usually involves minimizing a target cost function, which, in our case, 
is the proposed optimization target.
In the past, many papers have proposed to use convex optimization techniques 
to solve rate control related problems.
Y. Sermadevi \textit{et al.} \cite{sermadevi2006convex} proposed an iterative convex programming framework to optimize the rate control performance in VBR streaming under multiple channel rate constraints.
C. Pang \textit{et al.} \cite{pang2013dependent} proposed a convex optimization model to solve the joint bit allocation problem in H.264 statistical multiplexing, where the model has inter-frame dependency compatibility.
This paper intends to introduce the optimization methodology to LFI compression. 

\section{The Optimization Target} \label{sec:opt-target}

In this section, we introduce the weighted distortion model to measure the spatial quality affected by optical distortion in Section \ref{subsec:sq}.
Then, in Section \ref{subsec:ac}, angular consistency will be modeled by
a smoothness penalty term,
which discourages coding distortion from fluctuating wildly in the 4-D space.
Finally, in Section \ref{subsec:opt-target}, a generic optimization target that incorporates these two models in 
a natural and principled way will be proposed under the Bayesian maximum a posteriori (MAP) criteria.
\subsection{Spatial quality and weighted distortion} \label{subsec:sq}

The traditional PSNR metric is defined via the mean square error (MSE).
Given a lossless image $I$ and a noisy image $R$, both with resolution $M\times N$, the MSE is defined as
\begin{align} \label{eq:mse}
	\mli{MSE}=\frac{1}{MN}\sum_{i=1}^M\sum_{j=1}^N(I_{ij}-R_{ij})^2,
\end{align}
and the PSNR (in dB) is defined by
\begin{align} \label{eq:psnr}
	\mli{PSNR}=10\cdot \log_{10}\bigg(\frac{\mli{MAX}_I^2}{\mli{MSE}}\bigg),
\end{align}
where $\mli{MAX}_I$ is the maximum pixel value,
$I_{ij}$/$R_{ij}$ is the pixel value of $I$/$R$ at row $i$ and column $j$. 
If the image has three color channels (e.g. RGB), 
it is first converted into the YUV color space,
and the weighted average of the PSNR for each channel is computed,
with weights 6(Y):1(U):1(V) \cite{viola2017comparison}.

Since the LFI has four dimensions, 
the computation of its PSNR is often based on its SAIs, which have two dimensions as planar images.
The arithmetic mean of PSNRs of the SAIs have been used in literature as the objective metric:
\begin{align} \label{eq:psnr-viola}
	\mli{PSNR} = \frac{1}{(K-2)(L-2)}\sum_{k=2}^{K-1}\sum_{l=2}^{L-1}\mli{PSNR}(k,l),
\end{align}
where $\mli{PSNR}(k,l)$ is the PSNR of the SAI with $uv$ coordinates $(k,l)$,
$K$/$L$ is the resolution of the $u$/$v$ dimension.
There are two shortcomings of using the metric (\ref{eq:psnr-viola}).
Firstly, taking the arithmetic mean of PSNRs is equivalent to taking the geometric mean of MSEs, as (\ref{eq:mse})(\ref{eq:psnr}) and (\ref{eq:psnr-viola}) give
\begin{align} \label{eq:psnr-propto}
	\mli{PSNR} \propto -\log_{10} \displaystyle{\prod_{k=2}^{K-1}\prod_{l=2}^{L-1}\sum_{i=1}^M}\sum_{j=1}^N(I_{klij}-R_{klij})^2,
\end{align}
where $I_{klij}$/$R_{klij}$ is the pixel value of the LFI $I$/$R$ with $uvxy$ coordinates $(k,l,i,j)$.
The arithmetic mean is taken in the $xy$ dimensions while the geometric mean is taken in the $uv$ dimensions, therefore (\ref{eq:psnr-viola}) introduces an artificial asymmetry in the 4-D space.
Secondly, due to the optical distortion, the most external views are excluded in (\ref{eq:psnr-viola}).
However, some non-excluded external views may still be completely black, which can be perfectly encoded without distortion, yielding an overall PSNR of infinity by (\ref{eq:psnr-viola}).
An exploitative encoder can encode a single SAI in lossless mode to increase the overall PSNR without bound,
which is unfair to normal encoders.

\begin{figure}[!t]
\centering
\includegraphics[width=0.5\textwidth]{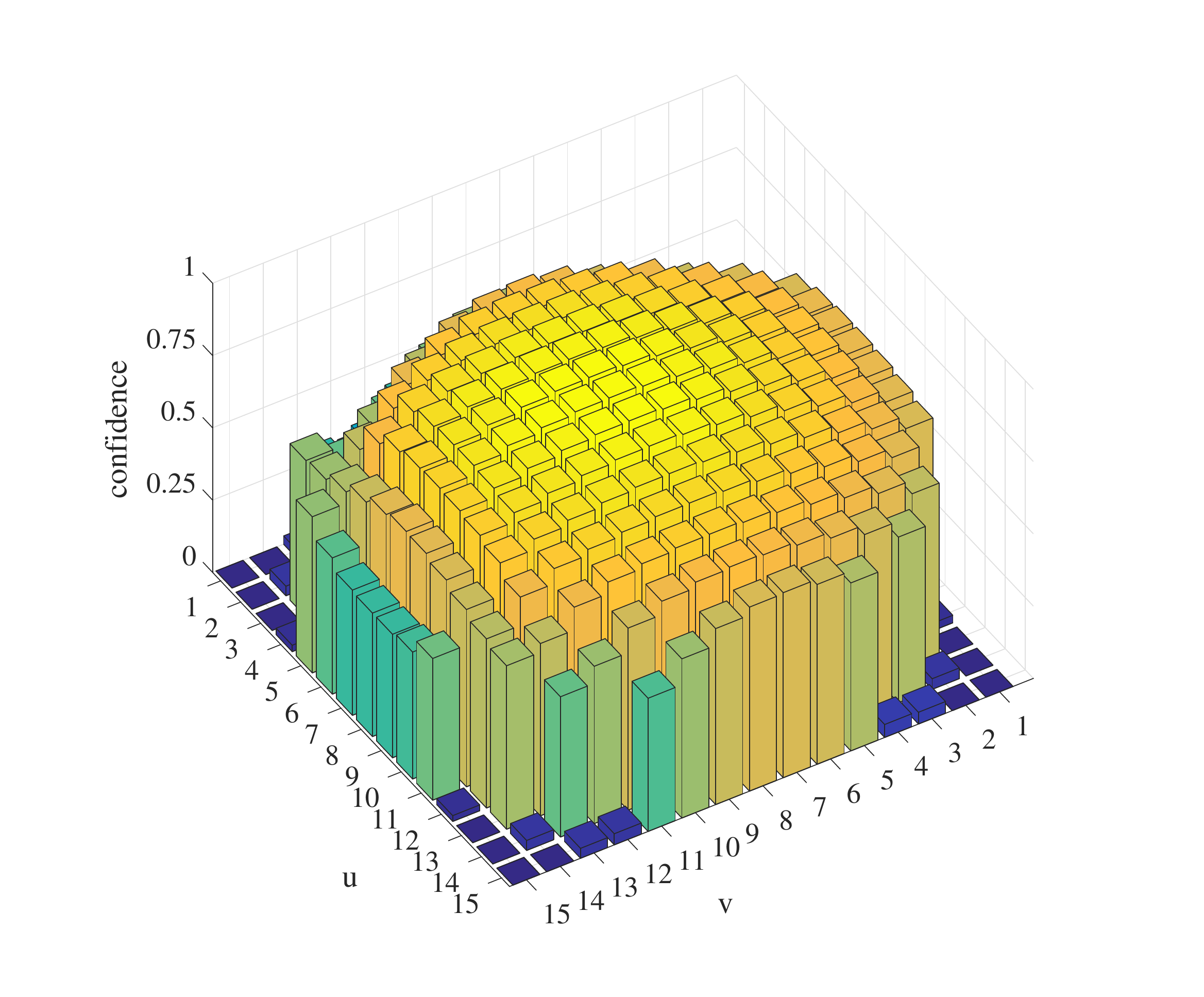}
\caption{An example of the distribution of pixel confidence $w_{kl}$ on the $uv$ plane, derived from the LFI \textit{Stone pillars outside} in dataset \cite{rerabek2016new}.}
\label{fig:confidence}
\end{figure}

We propose a weighted MSE to cope with the shortcomings mentioned above.
The optical distortion of captured LFIs results from 
the geometric property of light rays passing through external micro-lenses.
This optical distortion causes most external $uv$ views to lose fidelity.
The negative effect can be quantitatively measured
by an extra channel representing the confidence of each pixel.
The dataset in \cite{rerabek2016new} provides light field data with this extra channel.
Most pixels in the same SAI have similar levels of confidence,
since they are captured by the same micro-lense,
hence in this paper, $w_{kl}$ will be used to denote the average confidence of pixels in the SAI with $uv$ coordinates $(k,l)$.
A typical distribution of $w_{kl}$ on the $uv$ plane is drawn in Fig.~\ref{fig:confidence}.
It is consistent with the intuition that external SAIs have lower confidence.

As argued in Section \ref{sec:introduction}, the pixel confidence affects the amount of information lost due to compression.
 Suppose a captured LFI $I$ is compressed and decompressed,
 the decompression yields a reconstructed LFI $R$,
 with $I_{klij}$ and $R_{klij}$ defined as in (\ref{eq:psnr-propto}).
 If $w_{kl}$ is large, $I_{klij}$ has high fidelity and hence contains more reliable information about the ground-truth.
 The difference between  $I_{klij}$ and $R_{klij}$ represents the amount of information lost due to compression,
 and since $w_{kl}$ is large, more reliable information is lost comparing to pixels with lower confidences.
 Based on this analysis, we define a weighted MSE as
 \begin{align} \label{eq:weighted-mse}
 	\mli{wMSE} = \frac{1}{KLMN}\sum_{k=1}^{K}\sum_{l=1}^{L}\sum_{i=1}^{M}\sum_{j=1}^{N}\phi(w_{kl})(I_{klij}-R_{klij})^2,
 \end{align}
 where $\phi(w_{kl})$ is the weight assigned to the pixel at $uvxy$ coordinates $(k,l,i,j)$.
 Since $w_{kl}$ is shared by all pixels in the same SAI,
 if we use $\mli{MSE}(k,l)$ to denote the MSE of the SAI with $uv$ coordinates $(k,l)$, (\ref{eq:weighted-mse}) can be simplified as
 \begin{align} \label{eq:weighted-mse-simplified}
 	\mli{wMSE} = \frac{1}{KL}\sum_{k=1}^{K}\sum_{l=1}^{L}\phi(w_{kl})\cdot \mli{MSE}(k,l).
 \end{align}
 
 As argued previously, a larger $w_{kl}$ corresponds to a larger contribution to the loss, 
 hence the function $\phi$ is monotonically increasing.
 Practically $w_{kl}$ provided in the dataset may take values in a wide range, 
 and it is therefore convenient to rescale $w_{kl}$ to the unit interval $[0,1]$, 
 and define $\phi(\cdot)$ accordingly.
 A simple choice, used in our previous work \cite{guo2018convex}, is 
 \begin{align}
 	\phi(w) = w^2,
 \end{align}
where the confidence is squared to match the squared error in MSE.
The weighted MSE given by (\ref{eq:weighted-mse-simplified}) offers a symmetric treatment to the 4-D space, 
as the arithmetic mean is taken in all four dimensions.
Also, a single SAI compressed in lossless mode does not have a dominant impact on the weighted MSE.
Practically, the LFI contains three color channels.
To compute the weighted MSE, the LFI is firstly converted into the YUV color space,
and weights are assigned to each channel as 6(Y):1(U):1(V), in a similar way as PSNR:
\begin{align}
	\mli{wMSE} = \frac{6\cdot \mli{wMSE}_Y + \mli{wMSE}_U + \mli{wMSE}_V}{8}.
\end{align}

\subsection{Angular consistency and smoothness penalty} \label{subsec:ac}

The LFI contains complete information of the light rays in the scene,
therefore, a primary application of the LFI is to generate
new views with changed viewpoints or changed focal points.
This involves computing the intensity of the light ray from an arbitrary point $(u, v)$ on the $uv$ plane to another arbitrary point $(x,y)$ on the $xy$ plane.
Since the captured LFI (denoted by $I$) only contains a finite number of pixels,
an interpolation procedure is required.

Formally, suppose that only pixels with integer-valued coordinates are captured in $I$.  To obtain the pixel value at an arbitrary real-valued 4-D coordinates $(u, v, x, y)$,
captured pixels in $I$ that have coordinates close to $(u, v, x, y)$ can be used to interpolate the result.
M. Levoy \textit{et al.} \cite{levoy1996light} proposed to use the quadri-linear interpolation,
where the nearest two integer-positioned points in each dimension are selected,
giving a total of sixteen integer-positioned points as input:
\begin{align} \label{eq:quadrilinear}
	\tilde{I}_{uvxy}=\mli{Quadrilinear}(I_{0000},I_{0001},...,I_{1111}),
\end{align}
where $\tilde{I}_{uvxy}$ is the approximated pixel value at coordinates $(u,v,x,y)$, and $I_{klij}(0\le k,l,x,y \le 1)$ is defined as the pixel value at coordinates
$(\lfloor u \rfloor + k, \lfloor v \rfloor + l, \lfloor x \rfloor + i, \lfloor y \rfloor + j)$.
Here $\lfloor x \rfloor$ denotes the greatest integer less than or equal to $x$.

\begin{figure}[!t]
\centering
\includegraphics[width=0.4\textwidth]{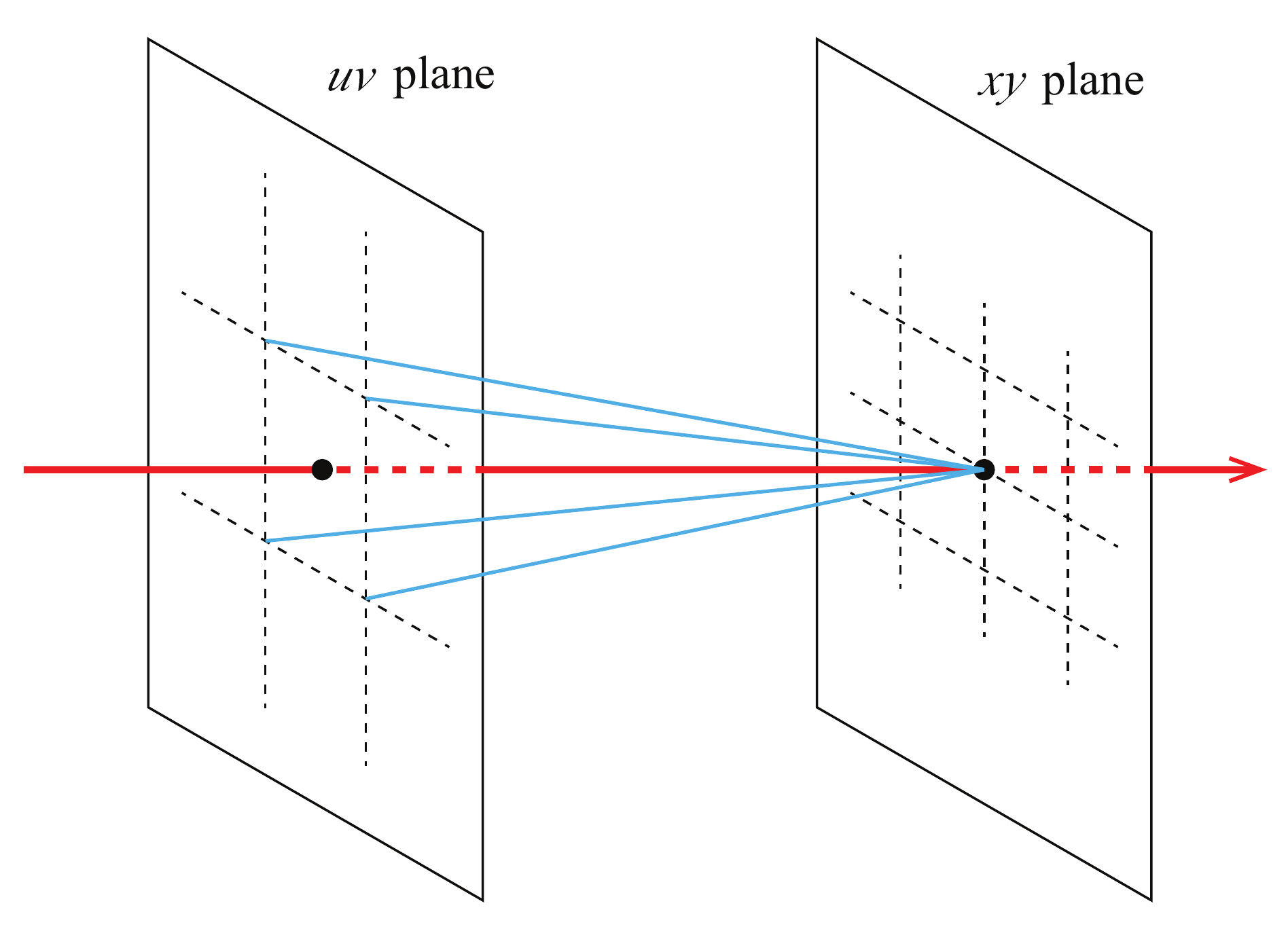}
\caption{Illustration of the bilinear interpolation on the $uv$ plane. The pixel with real-valued coordinates represented by the red line is interpolated by the nearest four pixels with integer-valued coordinates, represented by the blue lines.}
\label{fig:bilinear}
\end{figure}

In \cite{levoy1996light}, it is argued that if sampling rates of the $uv$ and $xy$ planes are different, 
the benefits of filtering each plane may be different. 
For a multi-lense array light field camera, 
the $uv$ resolution is determined by the multi-lense array, 
and the $xy$ resolution is determined by the image sensor.
Due to the manufacturing technology, the $xy$ resolution is usually much higher than the $uv$ resolution. 
For example, LFIs in the dataset \cite{rerabek2016new} captured by a Lytro Illum have an $xy$ resolution of 434$\times$625
and a $uv$ resolution of 15$\times$15.
In practical applications, the $xy$ resolution is usually large enough so that
 it is unnecessary to use real-valued $xy$ coordinates to achieve a higher precision.
Therefore, (\ref{eq:quadrilinear}) can be simplified to a bilinear interpolation only involving $uv$ coordinates:
\begin{align} \label{eq:bilinear}
	\tilde{I}_{uv} &= (\lceil u \rceil - u)(\lceil v \rceil - v)I_{00} + (\lceil u \rceil - u)(v - \lfloor v \rfloor)I_{01} \nonumber \\
	&+ (u - \lfloor u \rfloor)(\lceil v \rceil - v)I_{10} + (u - \lfloor u \rfloor)(v - \lfloor v \rfloor)I_{11},
\end{align}
where the $xy$ coordinates are omitted for clarity,
and $\lceil x \rceil$ denotes the least integer greater than or equal to $x$.
An illustration of the simplified bilinear interpolation on the $uv$ plane is shown in Fig.~\ref{fig:bilinear},
where the blue lines represent the pixel values $I_{00}$, $I_{01}$, $I_{10}$, and $I_{11}$,
the red line represents the interpolated value $\tilde{I}_{uv}$.

The coding distortion incurred during lossy compression can affect the interpolation result, according to (\ref{eq:bilinear}).
Denote the reconstructed light field image from decompression by $R$,
and define $\tilde{R}$, $R_{kl}$ similarly as $\tilde{I}$, $I_{kl}$.
The coding distortion is then measured by $D_{kl} = I_{kl} - R_{kl}$.
The interpolation gives a biased result due to the coding distortion:
\begin{align} \label{eq:I-R}
	&\tilde{I}_{uv} - \tilde{R}_{uv} \nonumber\\
	 =& (\lceil u \rceil - u)(\lceil v \rceil - v)D_{00} + (\lceil u \rceil - u)(v - \lfloor v \rfloor)D_{01} \nonumber \\
	+& (u - \lfloor u \rfloor)(\lceil v \rceil - v)D_{10} + (u - \lfloor u \rfloor)(v - \lfloor v \rfloor)D_{11}.
\end{align}

There are two reasons that adjacent pixels on the $uv$ plane should have similar amount of coding distortion.
Firstly, $D_{kl}$ represents the quantization residue during lossy compression
and is widely modeled with independent symmetric distributions (e.g. Cauchy/Laplace distribution) in the signal processing community \cite{altunbasak2004analysis}.
The positive distortion can therefore possibly offset the negative distortion in (\ref{eq:I-R}), if $\lvert D_{00} \rvert \approx \lvert D_{01} \rvert \approx \lvert D_{10} \rvert \approx \lvert D_{11} \rvert$.
In contrast, if the coding distortion is concentrated on one pixel, say,
$\lvert D_{00} \rvert \gg \lvert D_{01} \rvert \approx \lvert D_{10} \rvert \approx \lvert D_{11} \rvert$, 
the distortion cannot be offset and most pixels in the unit square $(\lfloor u \rfloor, \lceil u \rceil) \times 
(\lfloor v \rfloor, \lceil v \rceil)$ will have a large distortion with the same sign as $D_{00}$.
Secondly, the LFI can be used to generate videos with moving viewpoints. 
This can be done by fixing the $xy$ coordinates and moving on the $uv$ plane.
A large distortion difference of adjacent pixels on the $uv$ plane will cause
significant video quality fluctuations, which can negatively affect the overall quality of experience (QoE).

Unfortunately, in PTS-based compression, an LFI is decomposed into SAIs, which are arranged into a PTS according to some specific order.
There is no guarantee that adjacent SAIs on the $uv$ plane
will also be adjacent in the temporal order.
Therefore, wild distortion fluctuations may exist among adjacent frames in $uv$ coordinates.
To measure these distortion fluctuations,
we propose a smoothness penalty (SP) term based on the squared difference of MSE among adjacent SAIs:
\begin{align} \label{eq:sp}
	\mli{SP} = \sum_{\substack{k,l,m,n\\1\le k,m \le K\\1\le l,n \le L\\(k,l)\neq (m,n)}} \varphi(k,l,m,n)(\mli{MSE}(k,l) - \mli{MSE}(m,n))^2,
\end{align}
where $K,L$ are defined as in (\ref{eq:weighted-mse}),
$\mli{MSE}$ is defined as in (\ref{eq:weighted-mse-simplified}),
and $\varphi(k,l,m,n)$ describes the adjacency of the pair $(k,l)$ and $(m,n)$.
The pixel confidence $w_{kl}$ and $w_{mn}$ should also be taken into account,
as the smoothness penalty is based on the rendering quality,
and regions with higher pixel confidence are more prioritized.
Therefore, the function $\varphi(\cdot)$ is factored into two parts:
\begin{align} \label{eq:varphi}
	\varphi(k,l,m,n) = \delta((k,l),(m,n))W(w_{kl},w_{mn}).
\end{align}
By (\ref{eq:I-R}), adjacent pixels are at least 8-connected, therefore the function $\delta(\cdot)$ should be zero if $(k,l)$ and $(m,n)$ are not 8-neighbors. 
In this paper, the function $\delta(\cdot)$ is defined as
\begin{align} \label{eq:delta}
	\delta((k,l),(m,n)) = \left\{\begin{array}{ll}
                  			2 & \text{if~} \lvert k - m \rvert + \lvert l-n \rvert = 1,\\
                  			1 & \text{if~} \lvert k - m \rvert = \lvert l-n \rvert = 1,\\
                  			0 & \text{otherwise.}
                  			\end{array}
                  			\right.
\end{align}
In other words, 4-connected pairs take value 2, 
8-connected but not 4-connected pairs take value 1, 
and other pairs are excluded.
The reason for distinguishing 4-neighbors and 8-neighbors will be explained in Section \ref{subsec:opt-target}.
For filters other than bilinear interpolation, other forms of $\delta$ can be used.
The function $W$ is defined by
\begin{align} \label{eq:W}
	W(w_{kl},w_{mn}) = \phi(\min(w_{kl},w_{mn})),
\end{align}
which implies that pixel pairs are considered prioritized only if both pixels have high confidence.

\subsection{Optimization target} \label{subsec:opt-target}

With the weighted MSE and the smoothness penalty,
we can form an optimization target that incorporates both aspects, 
such that when the rate-distortion is optimized toward this target, 
both spatial quality and angular consistency will be addressed.
We will first show that the weighted MSE can be generalized to the continuous space.
As in Section \ref{subsec:ac}, it is assumed that the $xy$ resolution is high enough to avoid using continuous coordinates.
Fix the $xy$ coordinates, and consider the expected value of the squared error in the unit square $I^2=(\lfloor u \rfloor, \lceil u \rceil) \times 
(\lfloor v \rfloor, \lceil v \rceil)$ on the $uv$ plane:
\begin{align}
	\mathbb{E}[(\tilde{I}_{uv}-\tilde{R}_{uv})^2] = \int_{I^2}(\tilde{I}_{uv}-\tilde{R}_{uv})^2\d u\d v,
\end{align}
where $\tilde{I}_{uv}$ and $\tilde{R}_{uv}$ are defined as in (\ref{eq:I-R}).
This integral can be evaluated analytically with straightforward calculations:
\begin{align}  \label{eq:cond-expectation}
	&\mathbb{E}\bigg[\int_{I^2}(\tilde{I}_{uv}-\tilde{R}_{uv})^2\d u\d v~\bigg\vert~ D_{00},D_{01},D_{10},D_{11}\bigg] \nonumber \\ 
	=&\frac{1}{9}(D_{00}^2+D_{01}^2+D_{10}^2+D_{11}^2) + \frac{1}{18}(D_{00}D_{11}+D_{01}D_{10}) +\nonumber \\ 
	&\frac{1}{9}(D_{00}D_{01}+D_{00}D_{10}+D_{01}D_{11}+D_{10}D_{11}).
\end{align}
The first term is the finite sample sum of squared error,
the second and the third terms are adjustment terms for 8-neighbors and 4-neighbors.
Here we justify the different treatments to 8-neighbors and 4-neighbors in (\ref{eq:delta}),
as their coefficients are different in (\ref{eq:cond-expectation}).
As explained in Section \ref{subsec:ac},
$D_{kl}$s are uncorrelated and symmetric to the origin.
Take expectation to (\ref{eq:cond-expectation}) again to obtain
\begin{align}
	\mathbb{E}[(\tilde{I}_{uv}-\tilde{R}_{uv})^2\vert D_{00}^2+D_{01}^2+D_{10}^2+D_{11}^2=D] = \frac{D}{9}.
\end{align}
This conditional expectation implies that 
the expected value of squared error in the continuous space is proportional to
the finite sample MSE.
Therefore, we can use the weighted MSE (\ref{eq:weighted-mse-simplified}) as an unbiased evaluation for spatial quality, even if the changed viewpoint or focus point does not have integer coordinates.

To account for the distortion fluctuations on the $uv$ plane, a smoothness penalty term is added, whose strength is controlled by a coefficient $\lambda$.
The smoothness penalty is taken square root and averaged across all SAIs to match the weighted MSE (\ref{eq:weighted-mse}), so that they have the same magnitude:
\begin{align} \label{eq:T-metric}
	T = \mli{wMSE} + \lambda \frac{\sqrt{\mli{SP}}}{KL}.
\end{align}

The proposed optimization target (\ref{eq:T-metric}) can be interpreted from a probabilisitic view.
Define $G$ as the ground-truth light field, 
$I$ as the captured LFI with finite samples, 
and $R$ as the reconstructed LFI from decompression.
In this model, $G$ has unknown fixed values, $I$ is a random variable (due to the optical distortion) that contains information about $G$, and $R$ is a random variable (due to the coding distortion) as a function of $I$.
The quality of $R$ can be measured by $\mathbb{P}(G=R~\vert~ I)$,
the probability that the reconstruction is faithful to the ground truth,
given the captured samples.
The prior probability $\mathbb{P}(G=R)$ can be determined from angular consistency,
since a ground truth light field shall provide consistent visual quality in all perspectives.
This can be expressed mathematically as a Gibbsian form prior distribution \cite{winkler2012image}:
\begin{align} \label{eq:bayes-prior}
	\mathbb{P}(G=R) = Z_1^{-1} \exp(-\alpha \cdot \sqrt{\mli{SP}}),
\end{align}
where $Z_1$ is a normalization constant, $\alpha$ is a positive shape parameter. A large $SP$ term implies large inconsistency in visual qualities of different perspectives, which makes $G=R$ unlikely.
To model the evidence term $\mathbb{P}(I~\vert~ G=R)$,
consider the event $G=R$.
In this scenario, the weighted MSE (\ref{eq:weighted-mse}) becomes
 \begin{align} \label{eq:wmse-g}
 	\mli{wMSE} = \frac{1}{KLMN}\sum_{k=1}^{K}\sum_{l=1}^{L}\sum_{i=1}^{M}\sum_{j=1}^{N}\phi(w_{kl})(I_{klij}-G_{klij})^2,
 \end{align}
The optical distortion causes $I$ to deviate from $G$.
The squared error in (\ref{eq:wmse-g}) is weighted by pixel confidence
to reflect the fact that pixels with high confidence are less likely to deviate from ground-truth.
We therefore define the evidence term as
\begin{align} \label{eq:bayes-evidence}
	\mathbb{P}(I~\vert~ G=R) = Z_2^{-1}\exp(-\beta \cdot \mli{wMSE}),
\end{align}
where $Z_2$ and $\beta$ are the normalization constant and positive shape parameter as in (\ref{eq:bayes-prior}), respectively.
Combining (\ref{eq:bayes-prior}) and (\ref{eq:bayes-evidence}),
the posterior distribution satisfies
\begin{align}
	\mathbb{P}(G=R~\vert~ I) \propto \exp(-\mli{wMSE}-\alpha\beta^{-1}\sqrt{\mli{SP}}).
\end{align}
Therefore, if we set $\lambda=\alpha\beta^{-1}KL$, minimizing the target (\ref{eq:T-metric}) is equivalent to maximizing the posterior probability $\mathbb{P}(G=R~\vert~ I)$.

The selection of the strength factor $\lambda$ depends on the application under consideration.
For applications such as re-focused still image rendering,
the angular consistency is a minor factor and a small $\lambda$ should be chosen.
In contrast, the QoE of applications such as dynamic perspective video rendering heavily relies on the angular consistency, and a large $\lambda$ should be chosen.
Subjective evaluations can be conducted to aid the selection of $\lambda$.

\section{Frame level bit allocation framework} \label{sec:alloc}
In this section, frame level bit allocation for PTS-based
LFI compression will be formulated as an optimization problem in Section \ref{subsec:alloc-prob}.
The rate-distortion model for each PTS frame is proposed in Section \ref{subsec:alloc-rdm},
the proposed optimization problem will be solved with a two-step strategy
in Section \ref{subsec:alloc-solve}, and the two-pass encoding system will be described in Section \ref{subsec:alloc-enc}.
\subsection{Problem formulation} \label{subsec:alloc-prob}
As for the case of rate control for conventional image/video coding, rate control in PTS based LFI compression aims at achieving the best video quality while hitting a target bitrate. In PTS-based LFI compression, this can be done at the GOP, frame and CTU levels.
 In this paper, we will focus on bit allocation at the frame level, as it provides better granularity than the GOP level with a relatively lower complexity comparing to the CTU level.

Formally, denote the PTS frames in temporal order by $f_1, ..., f_n$.
Suppose the total bit budget is $R$,
and the bit cost of frame $f_i$ is $r_i$,
the frame level bit allocation can be formulated as an optimization problem:
\begin{align} \label{eq:opt}
\begin{split}
	\text{minimize } & T \\
	\text{s.t. } & \sum_{i=1}^n r_i \le R, r_i \ge 0,
\end{split}
\end{align}
where $T$ is the proposed optimization target defined by (\ref{eq:T-metric}).
If $T$ can be expressed as a function of $\bm r = (r_1,...,r_n)^\top$,
(\ref{eq:opt}) can be solved by searching for the best $\bm r$ in the $n$-simplex
$\lVert \bm r \rVert_1 \le R, \bm r \succcurlyeq \bm 0$.

In (\ref{eq:weighted-mse-simplified}), the weighted MSE is defined as a weighted average of the MSEs of each PTS frame.
Therefore, we can express the weighted MSE with $\bm r$ by expressing the MSE of $f_i$ with $\bm r$.
Similarly, since the SP term is also defined in terms of the MSE of $f_i$ by (\ref{eq:sp}), 
once the MSEs of each PTS frame are modeled by $\bm r$,
$T$ can be deduced by (\ref{eq:weighted-mse-simplified})(\ref{eq:sp}) and (\ref{eq:T-metric}).

\subsection{Frame level rate-distortion model} \label{subsec:alloc-rdm}
The previous subsection requires us to express the frame level MSE by the frame bit costs $\bm r$.
In predictive coding standards such as H.265/HEVC, 
the distortion of a frame depends not only on the bit cost of itself,
but also its reference frames.
Therefore, it is necessary to build different rate-distortion models for different reference frame structures.

\subsubsection{All-intra}
In the all-intra configuration, all PTS frames are encoded in intra mode (I-frame) and 
have no reference frames.
A PTS encoded with the all-intra configuration has a large bit cost,
but any SAI can be reconstructed without decoding other frames.
Therefore, the all-intra configuration is best suited for applications
that require fast SAI retrieval and have sufficient storage spaces.
In this case, the rate-distortion behavior of frame $f_i$ can be modeled by 
a convex hyperbolic function of $r_i$ as in \cite{mallat1998analysis}:
\begin{align} \label{eq:rd-intra}
	d_i = \alpha_i r_i^{\beta_i},
\end{align}
where $d_i$ is the MSE of $f_i$, and $\alpha_i>0$, $\beta_i<0$.
Ardestani \textit{et al.} showed in  \cite{ardestani2010rate} that the hyperbolic model (\ref{eq:rd-intra})
outperforms other empirical models such as the exponential model \cite{sullivan1998rate}.

\begin{figure}[!t]
\centering
\includegraphics[width=0.4\textwidth]{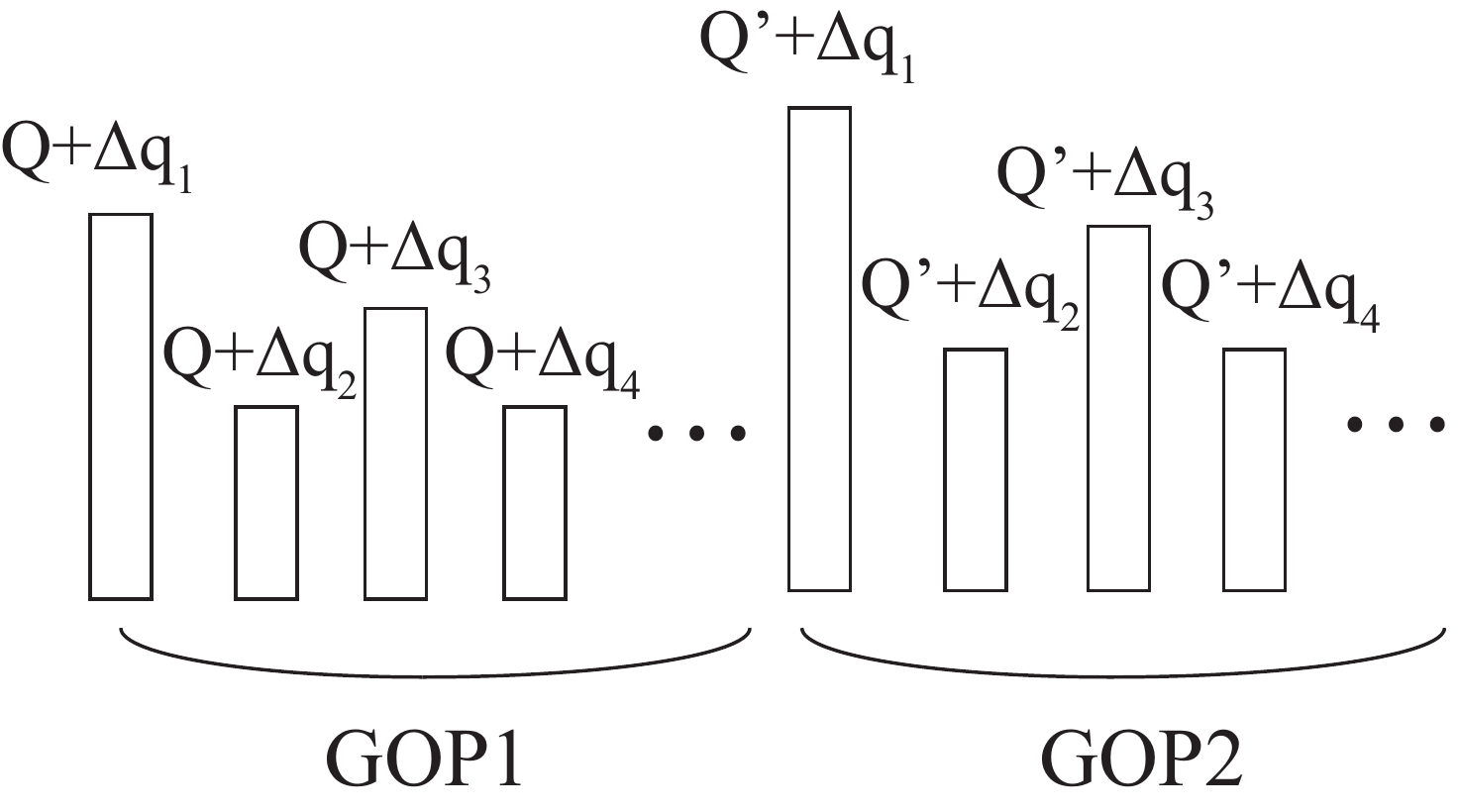}
\caption{Illustration of the QP offset structure of the hierarchical bit allocation scheme in HM 16.16 configurations. Each frame is assigned with a fixed QP offset $\Delta q_i$ depending on its position in the GOP. $Q$ and $Q'$ are the base QPs of GOP1 and GOP2, respectively.}
\label{fig:hier-qp}
\end{figure}

\subsubsection{Random-access}
In the random-access configuration,
I-frames are inserted at a fixed frequency
to provide access points,
and other frames are encoded as P/B-frames.
A PTS encoded with the random-access configuration can achieve a balance between
high compression rate and fast SAI retrieval,
since the P/B-frames are usually much smaller than I-frames, 
and any SAI can be reconstructed by simply decoding the corresponding GOP.
In this paper, we consider the configuration \textit{encoder\_randomaccess\_main\_GOP8.cfg} from HM 16.16, where
each GOP contains 8 frames, the first frame being an I-frame, and
a hierarchical bit allocation scheme is included to enhance the rate-distortion performance \cite{li2014lambda},
shown in Fig.~\ref{fig:hier-qp}.
For this configuration, since each GOP starts with an I-frame, we treat each GOP as an independent unit, 
such that the rate-distortion behavior of a frame has little dependence on frames in other GOPs.
Therefore, this paper proposes a novel approach that, given a fixed QP offset structure $\{\Delta q_i\}$ in the hierarchical bit allocation scheme,
the rate-distortion behavior of frame $f_i$ is modeled by a convex hyperbolic function of the total bit cost of the GOP containing $f_i$:
\begin{align} \label{eq:rd-ra}
	d_i = \alpha_i \bigg(\sum_{f_j\in G(i)}r_j\bigg)^{\beta_i},
\end{align}
where $G(i)$ denotes the GOP that $f_i$ belongs to.
This model is motivated by the fact that,
as long as the QP offset structure is fixed,
the bit cost of a frame, the bit cost of its reference frames, as well as
the total bit cost of the GOP all tend to move in the same direction,
which are negatively correlated with the frame distortion.

\subsubsection{Low-delay}
In the low-delay configuation, the first PTS frame is encoded as an I-frame,
and all subsequent frames are encoded as P/B-frames.
Since most frames are encoded with temporal prediction,
a PTS encoded with the low-delay configuration can achieve the highest compression efficiency,
but in order to retrieve a specific SAI,
all previous frames need to be decoded.
This configuration is best suited for applications with limited storage spaces  that allow long processing times.
In this paper, following our previous work \cite{guo2018convex}, we consider the configuration \textit{encoder\_lowdelay\_P\_main.cfg} from HM 16.16.
In theory, the rate-distortion behavior of a frame depends on all previous frames.
However, the dependence diminishes quickly as the temporal distance increases.
Inspired by this observation, this paper proposes a novel approach that,
by grouping the frames in the low-delay configuration into virtual GOPs, 
they can be treated as if they are encoded with the random-access configuration.
The virtual GOP size is chosen to be 12, and the previous argument is based
on the fact that, with a relatively large virtual GOP size, 
the rate-distortion dependency is large within the virtual GOP and small between virtual GOPs, 
as the average temporal distance between frames from different virtual GOPs 
are large.
Therefore, the rate-distortion model is also represented by (\ref{eq:rd-ra}),
where $G(i)$ denotes the virtual GOP that $f_i$ belongs to.

To estimate the parameters in (\ref{eq:rd-intra}) and (\ref{eq:rd-ra}),
trial compressions are conducted, which will be explained in detail
in Section \ref{subsec:alloc-enc}.
For the purpose of the next subsection,
the parameters are assumed to be determined 
and we are in a position to solve the optimization problem (\ref{eq:opt}).

\subsection{Solving the optimization problem} \label{subsec:alloc-solve}

In (\ref{eq:opt}), the target $T$ is defined by the MSE of PTS frames,
which are functions of frame bit costs $\bm r$, by (\ref{eq:rd-intra}) and (\ref{eq:rd-ra}).
The nature of the optimization problem (\ref{eq:opt}) depends on the functional form,
therefore, the way to solve (\ref{eq:opt}) depends on the configuration being used.
For clarity, the following notations are introduced.
For two PTS frames $f_i$ and $f_j$, whose corresponding $uv$ coordinates are 
$(k,l)$ and $(m,n)$, respectively, define
\begin{align}
	\phi(f_i) &= \phi(w_{kl}), \\
	\varphi(f_i, f_j) &= \varphi(k,l,m,n), \\
	d_i &= \mli{MSE}(k,l),
\end{align}
so that (\ref{eq:opt}) can be re-written in the form
\begin{align} \label{eq:opt-2}
	\begin{split}
	\text{minimize } & \sum_{i=1}^n \phi(f_i)d_i + \lambda \sqrt{\sum_{i,j}\varphi(f_i, f_j)(d_i-d_j)^2} \\
	\text{s.t. } & \sum_{i=1}^n r_i \le R, r_i \ge 0.
\end{split}
\end{align}

\subsubsection{All-intra}
The simplest case is the all-intra configuration where
$d_i$ is only related to $r_i$.
The domain of the problem is the $n$-simplex $\lVert \bm r \rVert_1 \le R, \bm r \succcurlyeq \bm 0$ in $\mathbb{R}^n$,
which is a convex set.
The composition rule of convexity implies that
replacing $d_i$ with a convex transform (\ref{eq:rd-intra}) of $\bm r$ in the linear weighted MSE term yields a convex function of $\bm r$,
but replacing $d_i$ with a non-linear transform (\ref{eq:rd-intra}) in the non-linear convex SP term may not yield a convex function.
Therefore, (\ref{eq:opt-2}) can not be directly solved with convex programming.
However, a two-step strategy can be used to find a good approximated solution:
\begin{enumerate}[label=(\alph*)]
	\item Solve (\ref{eq:opt-2}) by omitting the SP term to obtain an intermediate solution. Since the weighted MSE term is convex in $\bm r$, convex programming can be used.
	\item Approximate the SP term with a convex function of $\bm r$ using the intermediate solution, then (\ref{eq:opt-2}) can be solved with convex programming.
\end{enumerate}

\begin{figure}[!t]
\centering
\includegraphics[width=0.4\textwidth]{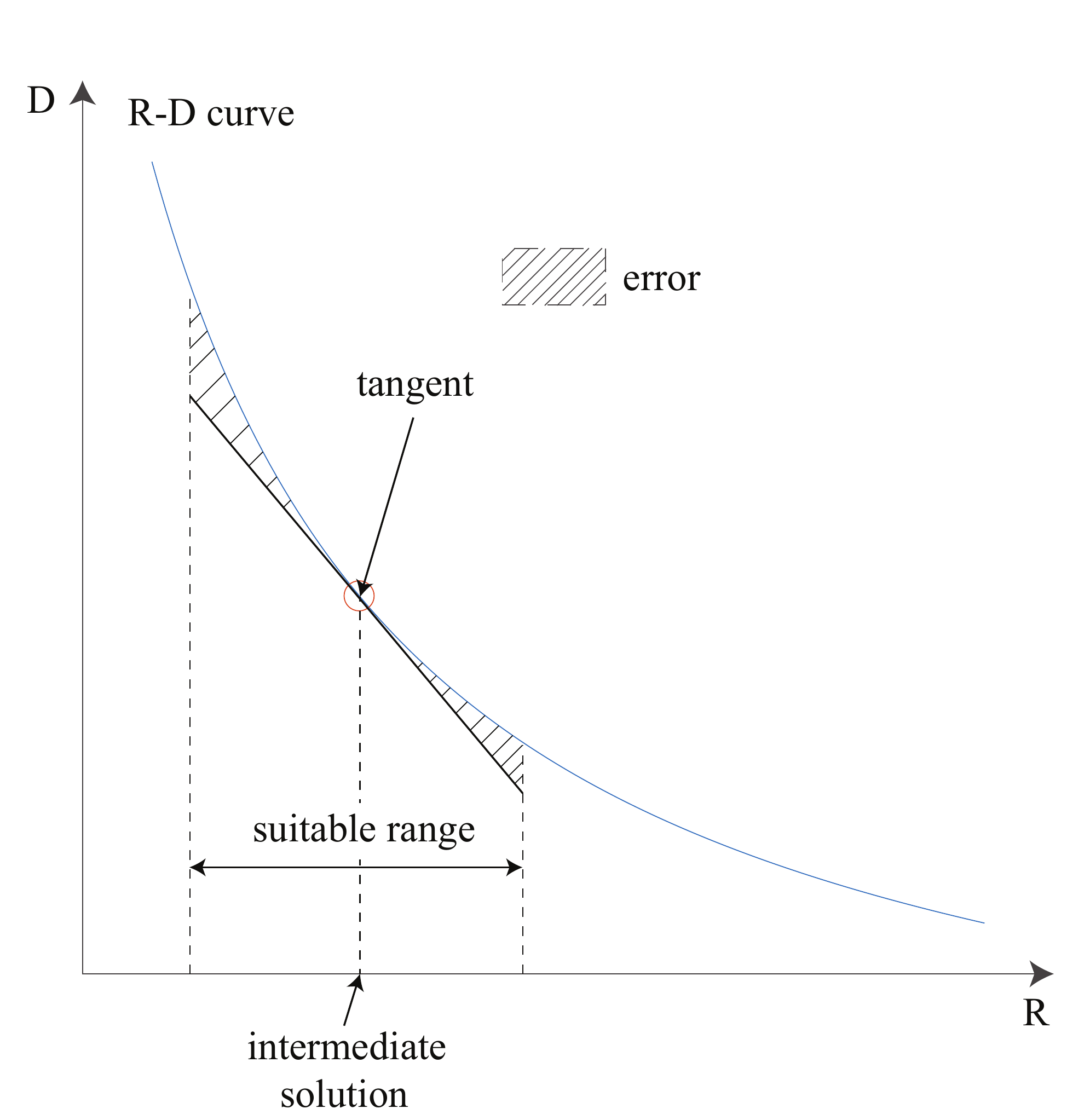}
\caption{Illustration of the first order approximation. The circle represents the obtained intermediate solution, as long as the true optimal solution is within the suitable range, the error caused by the first order approximation is limited.}
\label{fig:taylor}
\end{figure}

Suppose from step (a) we have obtained an intermediate solution $\mathcal{R}=(\mathcal{R}_1,...,\mathcal{R}_n)^\top$.
The rate-distortion model (\ref{eq:rd-intra}) is approximated by its first order Taylor expansion:
\begin{align} \label{eq:rd-intra-taylor}
	d_i = \alpha_i \beta_i \mathcal{R}_i^{\beta_i - 1}r_i + \alpha_i(1-\beta_i)\mathcal{R}_i^{\beta_i}.
\end{align}
This is a linear function of $r_i$, and plugging (\ref{eq:rd-intra-taylor}) into the convex SP term in (\ref{eq:opt-2}) yields a convex function of $\bm r$, which makes (\ref{eq:opt-2}) solvable with convex programming.
Note that in step (b), $d_i$ in the weighted MSE term is substituted by (\ref{eq:rd-intra}), 
and $d_i$, $d_j$ in the SP term are substituted by (\ref{eq:rd-intra-taylor}).

This two-step strategy relies on the assumption that the weighted MSE term
is more significant than the SP term, so that removing the SP term will
not cause a large deviation from the optimal solution,
which in turn makes (\ref{eq:rd-intra-taylor}) a good approximation.
An illustration of the first order approximation is shown in Fig.~\ref{fig:taylor}.
This assumption can be justified by the fact that a small weighted MSE always leads to a small SP term,
and subjectively speaking, a high fidelity is more prioritized than smooth transitions, especially in applications such as re-focused still image rendering.

Many convex programming packages \cite{cvx} require the target function to be expressed
in a regularized form.
This can be done by vectorization. 
Write $\bm d=(d_1,...,d_n)^\top$, it is straightforward to verify that
\begin{align}
\begin{split}
	&\sum_{i=1}^n \phi(f_i)d_i + \lambda \sqrt{\sum_{i,j}\varphi(f_i, f_j)(d_i-d_j)^2} \\
	=&\Phi^\top \text{pow}(\bm r,\bm \beta) + \lambda \cdot \sqrt{(Z\bm d)^\top \Psi (Z \bm d)} \\
	=&\Phi^\top \text{pow}(\bm r,\bm \beta) + \lambda \cdot \lVert \sqrt\Psi Z \bm d\rVert_2,
\end{split}
\end{align}
where $\Phi$ is a $n$-vector given by $\Phi_{i} = \phi(f_i)\alpha_i$,
$\text{pow}(\bm r,\bm \beta)$ is the element-wise power function $(r_1^{\beta_1},...,r_n^{\beta_n})^\top$.
$Z$ is an $n^2\times n$ difference coefficient matrix, each row corresponds to a pair of 
PTS frames. To be precise, row $k$ corresponds to $(f_i, f_j)$ where
\begin{align}
	i = \bigg\lceil \frac{k}{n} \bigg\rceil , j = k - n\bigg\lfloor \frac{k-1}{n} \bigg\rfloor, \label{eq:kij}\\
	Z_{kl} = \left\{\begin{array}{ll}
                  			1 & \text{if~} l=i,\\
                  			-1 & \text{if~} l=j,\\
                  			0 & \text{otherwise.}
                  			\end{array}
                  			\right. \label{eq:zkl}
\end{align}
$\Psi$ is an $n^2\times n^2$ diagonal matrix given by
\begin{align}\label{eq:psi} 
\Psi_{kk} = \varphi(f_i, f_j),
\end{align}
where $k, i, j$ are related by (\ref{eq:kij}).
To substitute $\bm r$ for $\bm d$, (\ref{eq:rd-intra-taylor}) gives
$\bm d = \bm c + B \bm r$, where $\bm c$ is an $n$-vector given by $c_i = \alpha_i(1-\beta_i)\mathcal{R}_i^{\beta_i}$, 
and $B$ is an $n\times n$ diagonal matrix given by $B_{ii} = \alpha_i \beta_i \mathcal{R}_i^{\beta_i-1}$. 
Combining all above gives
\begin{align}
	T = \Phi^\top \text{pow}(\bm r,\bm \beta) + \lambda \cdot \lVert \sqrt\Psi Z (\bm c + B \bm r) \rVert_2,
\end{align}
this is the vectorized form of the target function and is clearly convex in $\bm r$.
The optimization problem (\ref{eq:opt-2}) can then be solved with a convex programming package.

\subsubsection{Random-access and low-delay}
As explained in Section \ref{subsec:alloc-rdm}, the low-delay configuration is treated as the random-access configuration by grouping frames in the PTS into virtual GOPs,
with the GOP size set to 12 rather than 8.
Therefore, here we only need to consider a generic case,
where the GOP size is set to $k$.

According to (\ref{eq:rd-ra}), $d_i$ depends on $\bm r$ only through the
total bit costs of individual GOPs. 
Denote the GOPs by $G_1, ..., G_m$,
and the total bit costs of individual GOPs by
\begin{align}
	R_i = \sum_{f_j \in G_i} r_j,
\end{align}
the optimization problem (\ref{eq:opt-2}) can be transformed into an optimization problem with regard to $\bm R=(R_1, ..., R_m)^\top$:
\begin{align} \label{eq:opt-3}
	\begin{split}
	\text{minimize } & \sum_{i=1}^n \phi(f_i)d_i + \lambda \sqrt{\sum_{i,j}\varphi(f_i, f_j)(d_i-d_j)^2} \\
	\text{s.t. } & \sum_{i=1}^m R_i \le R, R_i \ge 0,
\end{split}
\end{align}
and the frame level rate-distortion model given by (\ref{eq:rd-ra}) is equivalent to
\begin{align} \label{eq:rd-ra2}
d_i = \alpha_i R_t^{\beta_i},
\end{align}
where $G_t=G(i)$, the (virtual) GOP that $f_i$ belongs to.

The optimization problem given by (\ref{eq:opt-3}) and (\ref{eq:rd-ra2})
can be solved with the same two-step strategy as the all-intra case.
The vectorized form of the target function is different due to the functional form.
If the intermediate solution is $\mathcal{R}=(\mathcal{R}_1,...,\mathcal{R}_m)^\top$, $T$ is given by
\begin{align}
	T = \sum_{i=1}^m \sum_{f_j \in G_i}\phi(f_j)\alpha_j R_i^{\beta_j} + \lambda \cdot \lVert \sqrt\Psi Z (\bm c + B \bm R) \rVert_2,
\end{align}
where $\Psi$ and $Z$ are given by (\ref{eq:kij})(\ref{eq:zkl})(\ref{eq:psi}) ,
while $\bm c$ is an $n$-vector given by
\begin{align}
	c_i = \alpha_i (1-\beta_i) \mathcal{R}_t^{\beta_i},
\end{align}
where $G_t=G(i)$, and $B$ is an $n\times m$ matrix given by
\begin{align}
	B_{ij} = \left\{\begin{array}{ll}
                  			\alpha_i \beta_i \mathcal{R}_{j}^{\beta_i -1} & \text{if~} G_j=G(i),\\
                  			0 & \text{otherwise.}
                  			\end{array}
                  			\right. \label{eq:bij}
\end{align}

\begin{figure}[!t]
\centering
\includegraphics[width=0.45\textwidth]{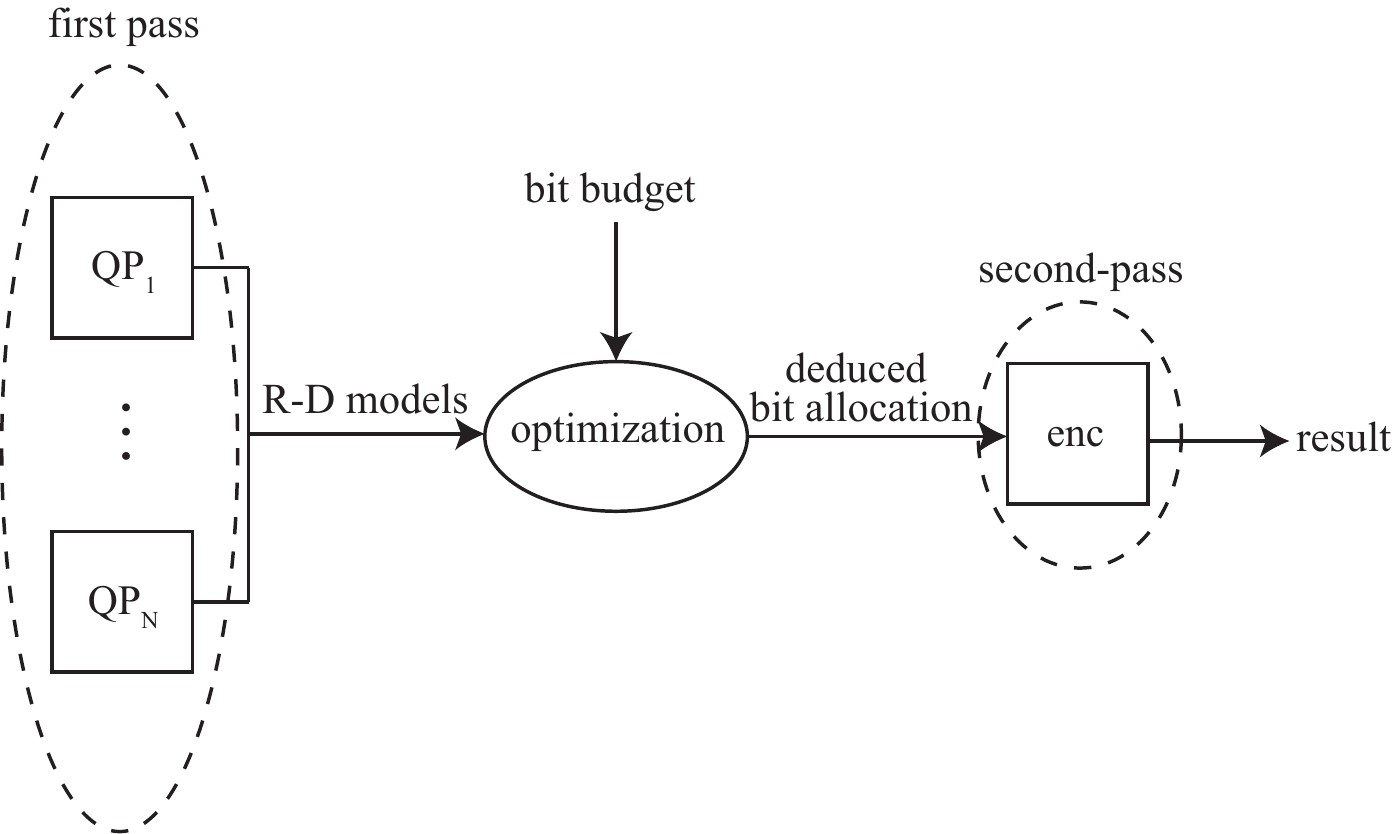}
\caption{Illustration of the two-pass encoding system. The exact forms of R-D models and the optimization depend on the encoder configuration.}
\label{fig:twopass}
\end{figure}

\subsection{Two-pass encoding system} \label{subsec:alloc-enc}
The proposed two-pass encoding system estimates the parameters $\alpha_i$ and $\beta_i$ in the rate-distortion model (\ref{eq:rd-intra}) or (\ref{eq:rd-ra}) during the first pass,
and the actual encoding is performed during the second pass according to the optimized solution of (\ref{eq:opt-2}) or (\ref{eq:opt-3}),
as shown in Fig.~\ref{fig:twopass}.
Depending on the encoder configuration used, different procedures are carried out.
\subsubsection{All-intra}
During the first pass, $N$ encoders $E_1, ..., E_N$ are run in parallel,
each encoder encodes the PTS frames with a constant QP,
e.g. all frames in $E_i$ are encoded with QP $q_i$.
Since in the all-intra configuration, the rate-distortion behavior of PTS frames is mutually independent,
the rate-distortion statistics can be obtained by collecting from all $N$ encoders.
For instance, suppose for frame $f_j$ the frame bit costs $r_j$ and MSEs $d_j$ collected
from all $N$ encoders are given by
\begin{align} \label{eq:rd-stats}
	\begin{split}
		r_j = r_{ji}, d_j = d_{ji}, \mli{QP} = q_i \text{ in } E_i (i=1,..., N).
	\end{split}
\end{align}
The rate-distortion model (\ref{eq:rd-intra}) can be estimated by regressing $\log d_j$ against $\log r_j$, using the data $d_{j1}, ..., d_{jN}$ and  $r_{j1}, ..., r_{jN}$:
\begin{align} \label{eq:reg-intra}
	\log d_j = \hat a_j + \hat b_j \log r_j,
\end{align}
the model parameters $\alpha_j$ and $\beta_j$ are given by
\begin{align} \label{eq:alpha-beta}
	\alpha_j = \exp \hat a_j, \beta_j = \hat b_j.
\end{align}

Once the model parameters are obtained, the optimal bit allocation $\tilde r=(\tilde r_1,...,\tilde r_n)^\top$ can be derived as in Section \ref{subsec:alloc-solve} by solving (\ref{eq:opt-2}).
During the second pass, frame QPs are selected to achieve minimum
deviation from the optimal bit allocation $\tilde r$.
For instance, suppose the rate-distortion statistics are again given by (\ref{eq:rd-stats}).
The QP for frame $f_j$ during the second pass is then given by $q_{j^\ast}$
where
\begin{align} \label{eq:abs-diffr1}
	j^\ast = \argmin_i |r_{ji} - \tilde r_j|.
\end{align}

\subsubsection{Random-access}
As explained in Section \ref{subsec:alloc-rdm}, the QP offset structure of the hierarchical bit allocation scheme is fixed and the optimization (\ref{eq:opt-3}) is performed at the GOP level.
To achieve this, we assign a base QP $q_i$ to each GOP $G_i$,
and the QP of frame $f_j$ is obtained by adding a QP offset according to its position in the GOP.
For instance, suppose the GOP size is $k$ and the QP offsets are given by $\Delta q_1, ...,  \Delta q_k$.
Frame $f_j$ is the $l$-th frame of GOP $G_t = G(j)$,
the QP of $f_j$ is then given by $q_t + \Delta q_l$.

During the first pass, $N$ encoders $E_1, ..., E_N$ are run in parallel,
each encoder encodes the PTS frames with a constant base QP,
e.g. all GOPs in $E_i$ are encoded with base QP $q_i$.
Since the rate-distortion behavior of different GOPs is assumed to be mutually independent,
again the rate-distortion statistics can be obtained by collecting from all $N$ encoders.
Using the same notations as above, for frame $f_j$ the frame bit costs $r_j$ and MSEs $d_j$ collected
from all $N$ encoders are given by
\begin{align} \label{eq:rd-stats2}
	\begin{split}
		r_j &= r_{ji}, d_j = d_{ji}, \\
		\mli{QP} &= q_i + \Delta q_l \text{ in } E_i (i=1,..., N).
	\end{split}
\end{align}
The rate-distortion model (\ref{eq:rd-ra2}) can be estimated by regressing 
$\log d_j$ against $\log R_{t}$, using the data $d_{j1}, ..., d_{jN}$ and $r_{j1}, ..., r_{jN}$:
\begin{align} \label{eq:reg-ra}
	\begin{split}
	\log d_j &= \hat a_j + \hat b_j \log R_{t}, \\
	R_{t} &= \sum_{f_s \in G_t} r_s.
	\end{split}
\end{align}
The model parameters are again given by (\ref{eq:alpha-beta}).
If (\ref{eq:opt-3}) yields the optimal bit allocation $\tilde r = (\tilde r_1, ..., \tilde r_m)^\top$,
the base QPs during the second pass are selected similarly by the minimum deviation criteria,
therefore the QP for $f_j$ is given by $q_{i^\ast} + \Delta q_l$ where
\begin{align} \label{eq:abs-diffr2}
	i^\ast = \argmin_i \bigg|\sum_{f_s \in G_t}r_{si} - \tilde r_t\bigg|.
\end{align}

\subsubsection{Low-delay}
As explained in Section \ref{subsec:alloc-rdm}, the low-delay configuration is treated as 
the random-access configuration, except that GOPs are replaced by virtual GOPs, 
and the GOP size $k$ is set to 12 instead of 8.

\section{Experimental Results} \label{sec:exp}
In this section, experimental results are summarized and explained 
to demonstrate the effectiveness of our proposed methods.
The experiment data preparation procedures are explained in Section \ref{subsec:exp-data}, 
the encoder specifications are given in Section \ref{subsec:exp-enc}.
The accuracy of the rate-distortion models is analyzed in Section \ref{subsec:exp-rdm},
and the performance of the proposed two-pass encoding system is shown in Section \ref{subsec:exp-sys}.

\subsection{Data preparation} \label{subsec:exp-data}
We selected 13 LFIs from the EPFL light-field dataset \cite{rerabek2016new},
representing different categories, shown in Table \ref{table_lfdata}.
The 4D LF MAT files contain four dimensional light field images, 
each has a resolution of 15$\times$15 for the $uv$ plane, and 434$\times$625 for the $xy$ plane.
They are firstly converted into individual SAIs, 
saved as 434$\times$625 8-bit png images. The png images are cropped to 432$\times$624 such that its resolutions are multiples of 4, and then arranged into pseudo-temporal sequences using the circular scan order proposed in \cite{hariharan2017low},
which are then converted into the YUV format.
Frames at the end of the PTS arranged according to the circular scan order
are black due to the severe distortion in most external SAIs.
These black frames are discarded and the YUV files only contain the first 192 frames.

\begin{table}[!t]
\renewcommand{\arraystretch}{1.3}
\caption{Selected Light-field Images}
\label{table_lfdata}
\centering
\begin{tabular}{|c|c|c|}
\hline
Category & LFI & Name\\
\hline
Buildings & I01 & Red and white building\\ \hline
Grids & I02 & Chain link fence 1\\ \hline
Landscapes & I03 & Reeds \\ \hline
Light & I04 & Graffiti \\ \hline
\multirow{2}{*}{Mirrors and transparency} & I05 & Broken mirror \\ \cline{2-3}
& I06 & Water drops \\ \hline
\multirow{2}{*}{Nature} & I07 & Caution bees \\ \cline{2-3}
& I08 & Swans 1 \\ \hline
\multirow{2}{*}{People} & I09 & Sphynx \\ \cline{2-3}
& I10 & Yan and Krios 1 \\ \hline
Studio & I11 & Desktop \\ \hline
\multirow{2}{*}{Urban} & I12 & Concrete cubes \\ \cline{2-3}
& I13 & Stone pillars outside \\ \hline
\end{tabular}
\end{table}

The 4D LF MAT files also contain the confidence of each pixel. 
They are firstly rescaled into the range $[0, 1]$, and the average confidence of each SAI are computed and stored for further processing.

\subsection{Encoder specifications} \label{subsec:exp-enc}
The encoder used in this paper is the HEVC reference software HM 16.16. Different specifications are given to the encoder
according to the configuration being used.
\subsubsection{All-intra}
In the all-intra configuration, all frames are encoded in the intra mode.
The detailed specifications are given by the default configuration file 
\textit{encoder\_intra\_main.cfg} from HM 16.16.
Four different total bit budgets are selected: 5Mb, 10Mb, 20Mb and 40Mb.
\subsubsection{Random-access}
In the random-access configuration, the intra period and the GOP size are both set to 8.
Other detailed specifications are the same as the default configuration file
\textit{encoder\_randomaccess\_main\_GOP8.cfg} from HM 16.16,
where the QP offsets are set to 1, 2, 3, 4, 4, 3, 4, 4.
Four different total bit budgets are selected: 1Mb, 2Mb, 4Mb and 8Mb.
\subsubsection{Low-delay}
In the low-delay configuration, only the first frame is encoded in the intra mode.
The detailed specifications are given by the default configuration file 
\textit{encoder\_lowdelay\_P\_main.cfg} from HM 16.16,
where the QP offsets are set to 5, 4, 5, 1.
Note that in our algorithm the virtual GOP size is 12,
 therefore for the purpose of Section \ref{subsec:alloc-enc}, 
 the QP offsets are repeated for two more times.
 Four different total bit budgets are selected: 500Kb, 1Mb, 2Mb and 4Mb.
 
\subsection{Rate-distortion models} \label{subsec:exp-rdm}
To estimate the rate-distortion models (\ref{eq:rd-intra})(\ref{eq:rd-ra2}),
30 encoders are run in parallel, with constant frame QP (or GOP base QP) ranging from 16 to 45.
The regressions (\ref{eq:reg-intra})(\ref{eq:reg-ra}) do not use all the available data,
 instead, in order to improve the prediction accuracy, 
 data from encoders yielding output sizes that are too large or small comparing to the target bit budget are discarded.
 In our experiments, a central QP $q_c$ is selected as the QP that gives the closest output size comparing to the target bit budget,
 and the data are selected for regression if they are encoded with QP from $\max(16, q_c - 7)$ to $\min(45, q_c + 7)$.
 
 \subsubsection{All-intra}
\begin{figure}[!t]

\begin{minipage}{.5\linewidth}
\centering
\subfloat[]{\label{main:a}\includegraphics[width=0.9\textwidth]{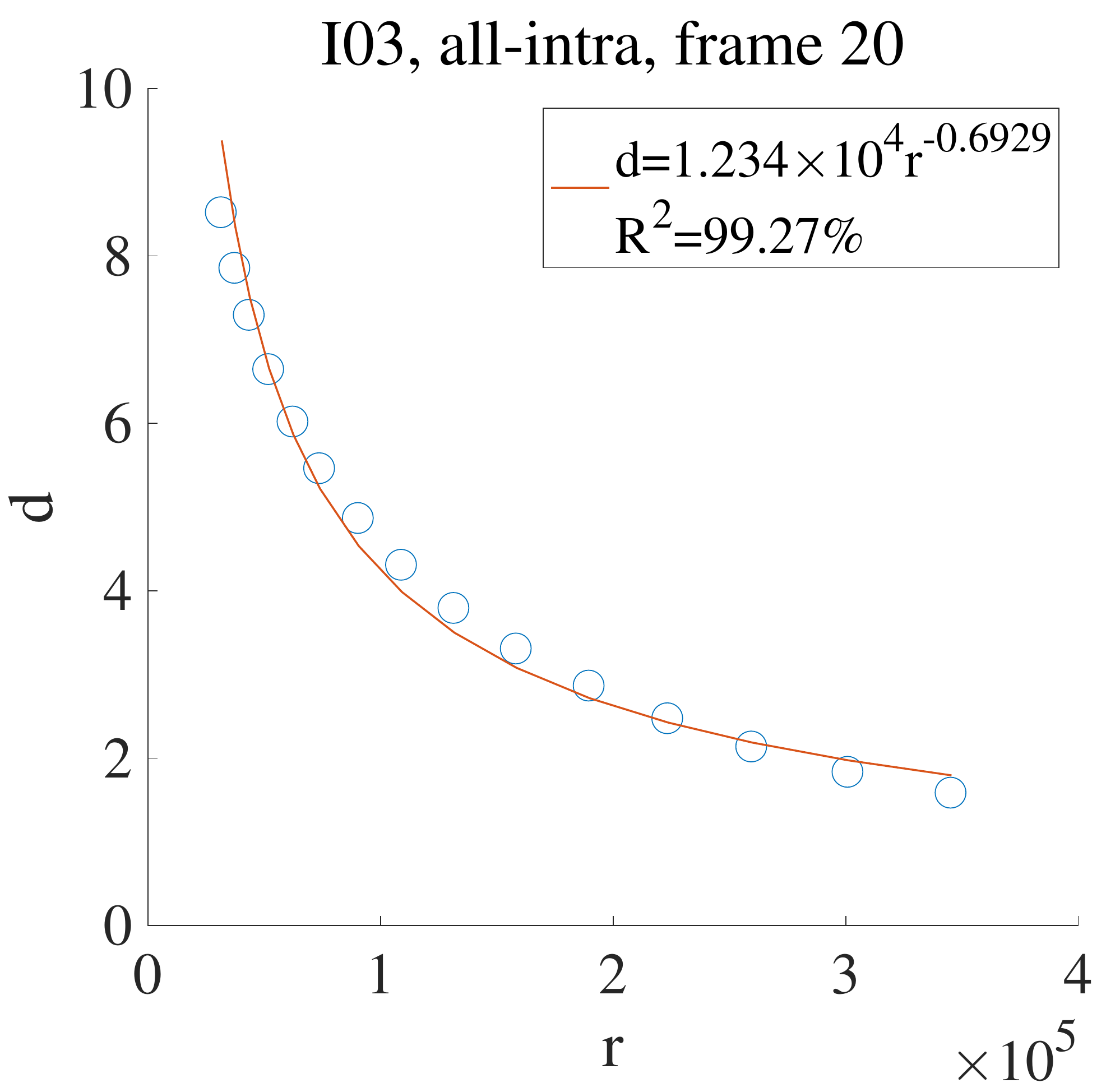}}
\end{minipage}%
\begin{minipage}{.5\linewidth}
\centering
\subfloat[]{\label{main:b}\includegraphics[width=0.9\textwidth]{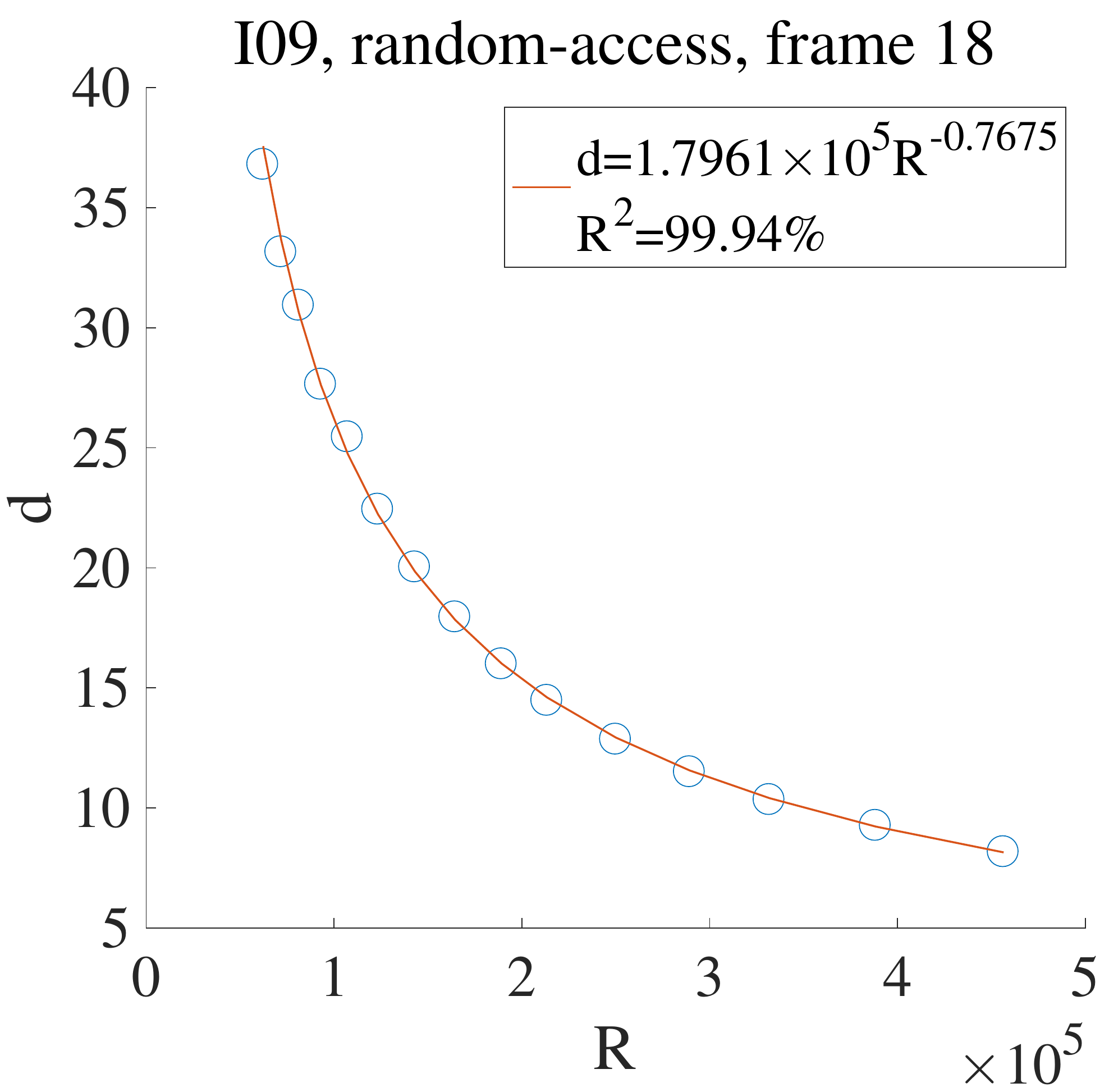}}
\end{minipage}\par\medskip
\centering
\subfloat[]{\label{main:c}\includegraphics[width=0.225\textwidth]{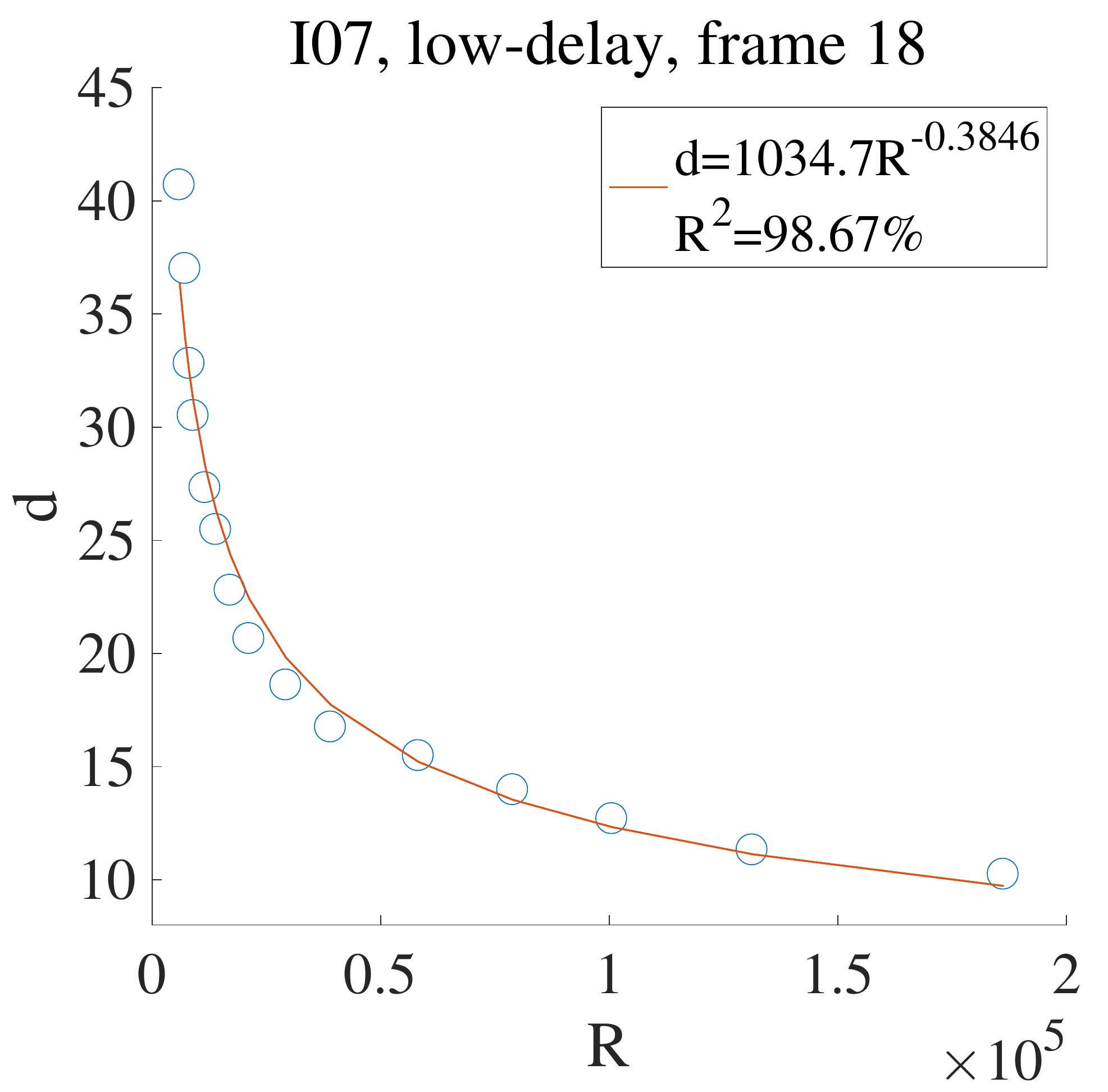}}

\caption{Examples of R-D model fitting results.}
\label{fig:rdm}
\end{figure}

 It is well known that the model (\ref{eq:rd-intra}) gives
 an accurate description of the frame rate-distortion behavior,
 holding all other factors fixed.
 The overall results are shown in Table \ref{table_rdresults} in the row \textit{all-intra}.
 The rate-distortion statistics of the 20th frames of all 13 LFIs
 are regressed using (\ref{eq:reg-intra}),
 where the target bit budget is set to 20Mb.
 The mean value and the standard deviation of the R-squared values
 are 99.39\% and 0.80\%, respectively,
 indicating a good performance of (\ref{eq:rd-intra}).
 The regression result of frame 20 in I03 is shown in Fig.~\ref{fig:rdm}(a) as an example.
 
\begin{table}[!t]
\renewcommand{\arraystretch}{1.3}
\caption{Experimental Results on Rate-distortion Models Fitting}
\label{table_rdresults}
\centering
\begin{tabular}{|c|c|c|c|c|}
\hline
Configuration & Frame & Bit budget & mean of $R^2$ & std of $R^2$\\
\hline
all-intra & 20 & 20Mb & 99.39\% & 0.80\%\\
\hline
\multirow{ 3}{*}{random-access} & 18 & 4Mb & 99.68\% & 0.33\%\\
 & 21 & 4Mb & 99.86\% & 0.11\%\\
 & 23 & 4Mb & 99.87\% & 0.12\%\\
\hline
\multirow{ 3}{*}{low-delay} & 14 & 2Mb & 99.58\% & 0.43\%\\
 & 18 & 2Mb & 99.61\% & 0.44\%\\
 & 24 & 2Mb & 99.55\% & 0.41\%\\
\hline
\end{tabular}
\end{table}

 \subsubsection{Random-access}
 For the proposed rate-distortion model (\ref{eq:rd-ra2}),
 its performance is tested against different frame positions in the GOP.
 The overall results are shown in Table \ref{table_rdresults} in the row \textit{random-access}.
 The rate-distortion statistics of 3 different frames in the 3rd GOP (frame 18, frame 21, and frame 23) of all 13 LFIs
 are regressed using (\ref{eq:reg-ra}).
 The target bit budget is set to 4Mb.
 It can be seen that although the regression performance varies with
 the position of the frame in its GOP, 
 in most cases (within two standard deviations) the R-squared's from the regressions are higher than 99\%.
 The regression result of frame 18 in I09 is shown in Fig.~\ref{fig:rdm}(b) as an example.
 
  \subsubsection{Low-delay}
 Similar to the random-access configuration,
 the performance of (\ref{eq:rd-ra2}) is also tested with various frame positions in the virtual GOP.
 The overall results are shown in Table \ref{table_rdresults} in the row \textit{low-delay}.
The rate-distortion statistics of 3 different frames in the 2nd virtual GOP (frame 14, frame 18 and frame 24) of all 13 LFIs
 are regressed using (\ref{eq:reg-ra}).
  The target bit budget is set to 2Mb.
 It can be seen that most R-squared are higher than 98\%.
  The regression result of frame 18 in I07 is shown in Fig.~\ref{fig:rdm}(c) as an example.

 \subsection{Two-pass encoding system} \label{subsec:exp-sys}
 
  \begin{table*}[!t]
\renewcommand{\arraystretch}{1.3}
\caption{Experimental Results on Rate-distortion Performances, BD-rates of All Cases}
\label{table_bdrates}
\centering
\resizebox{\textwidth}{!}{
\begin{tabular}{|c|c|c|c|c|c|c|c|c|c|c|c|c|}
\hline
\multirow{2}{*}{LFI} & \multicolumn{3}{c|}{all-intra} & \multicolumn{3}{c|}{random-access}  & \multicolumn{3}{c|}{low-delay} & \multicolumn{3}{c|}{low-delay, algorithm\cite{guo2018convex}} \\ \cline{2-13}
& $\lambda$=0 & $\lambda$=2 & $\lambda$=4 & $\lambda$=0 & $\lambda$=2 & $\lambda$=4 & $\lambda$=0 & $\lambda$=2 & $\lambda$=4 & $\lambda$=0 & $\lambda$=2 & $\lambda$=4 \\ \hline

I01 & -13.96\% & -11.65\% & -10.37\% & -14.29\% & -23.95\% & -31.54\% & -16.82\% & -18.59\% & -19.59\% & -4.16\% & -5.26\% & -6.29\%\\ \hline
I02 & -7.41\%  & -5.63\%  & -5.26\%  & -18.37\% & -29.89\% & -38.09\% & -0.69\%  & -2.90\%  & -4.63\%  & -3.86\% & -9.29\% & -14.56\%\\ \hline
I03 & -14.70\% & -13.97\% & -15.81\% & -12.43\% & -14.48\% & -15.20\% & -31.47\% & -33.22\% & -36.37\% & -11.52\%& -36.10\% & -46.65\%\\ \hline
I04 & -9.40\%  & -10.62\% & -12.44\% & -23.88\% & -36.05\% & -43.13\% & -16.70\% & -17.70\% & -18.23\% & -3.85\% & -4.74\%  & -5.63\% \\ \hline
I05 & -10.26\% & -10.08\% & -9.57\%  & -8.49\%  & -13.69\% & -17.75\% & -6.32\%  & -9.61\%  & -12.39\% & -6.35\% & -14.87\% & -21.26\%\\ \hline
I06 & -11.03\% & -13.80\% & -16.65\% & -18.84\% & -32.41\% & -44.46\% & -13.19\% & -15.29\% & -16.00\% & -5.69\% & -8.63\%  & -11.52\%\\ \hline
I07 & -10.91\% & -11.57\% & -13.05\% & -14.31\% & -25.19\% & -33.53\% & -5.15\%  & -7.54\%  & -9.77\%  & -1.14\% & -1.52\%  & -7.72\%\\ \hline
I08 & -8.87\%  & -9.27\%  & -10.47\% & -12.72\% & -21.83\% & -28.56\% & -16.47\% & -19.59\% & -22.67\% & -0.57\% & -2.25\%  & -3.80\%\\ \hline
I09 & -17.14\% & -16.76\% & -18.04\% & -15.74\% & -24.89\% & -32.36\% & -17.79\% & -20.51\% & -21.03\% & -3.21\% & -4.02\%  & -4.79\%\\ \hline
I10 & -8.13\%  & -7.95\%  & -8.91\%  & -14.03\% & -22.49\% & -29.06\% & -0.66\%  & -3.73\%  & -6.04\%  & -2.13\% & -10.67\% & -17.70\%\\ \hline
I11 & -15.63\% & -11.37\% & -8.02\%  & -15.25\% & -25.67\% & -34.20\% & -13.98\% & -12.90\% & -13.24\% & 1.39\% & -5.83\%  & -1.16\%\\ \hline
I12 & -13.34\% & -12.37\% & -12.45\% & -20.31\% & -32.63\% & -42.09\% & -3.51\%  & -3.97\%  & -4.64\%  & 0.11\%  & -0.64\%  & -1.34\%\\ \hline
I13 & -8.78\%  & -11.22\% & -13.96\% & -16.77\% & -26.91\% & -35.05\% & -13.83\% & -14.58\% & -19.41\% & -7.46\% & -11.67\% & -21.15\%\\ \hline
Average & -11.51\% & -11.25\% & -11.92\% & -15.80\% & -25.39\% & -32.69\% & -12.05\% & -13.86\% & -15.69\% & -3.73\% & -8.88\% & -12.58\%\\ \hline
\end{tabular}
}
\end{table*}

 \begin{table}[!t]
\renewcommand{\arraystretch}{1.3}
\caption{Experimental Results on Rate-distortion Performances, All-intra, $\lambda$ = 0 (Bits in Mbits)}
\label{table_rdperform}
\centering
\resizebox{0.48\textwidth}{!}{
\begin{tabular}{|c|c|c|c|c|c|c|}
\hline
\multirow{2}{*}{LFI} & \multirow{2}{*}{Bit budget} & \multicolumn{2}{c|}{proposed} & \multicolumn{2}{c|}{HM-RC} & \multirow{2}{*}{BD-rate}\\
\cline{3-6}
& & Bits & $T'$ & Bits & $T'$ & \\
\hline
\multirow{4}{*}{I01} & 5Mb & 5.067 & 36.04 & 5.023 & 35.47 & \multirow{4}{*}{-13.96\%}\\ \cline{2-6}
& 10Mb & 9.924 & 38.78 & 10.023 & 38.20 &\\ \cline{2-6}
& 20Mb & 20.002 & 41.87 & 20.021 & 41.15 &\\ \cline{2-6}
& 40Mb & 39.952 & 45.35 & 40.024 & 44.63 &\\
\hline
\multirow{4}{*}{I03} & 5Mb & 5.006 & 42.17 & 5.023 & 41.89 &\multirow{4}{*}{-14.70\%}\\ \cline{2-6}
& 10Mb & 9.994 & 43.68 & 10.024 & 43.38 &\\ \cline{2-6}
& 20Mb & 19.923 & 45.51 & 20.023 & 45.04 &\\ \cline{2-6}
& 40Mb & 39.834 & 47.96 & 40.021 & 47.26 &\\
\hline
\multirow{4}{*}{I07} & 5Mb & 5.056 & 33.90 & 5.023 & 33.51 &\multirow{4}{*}{-10.91\%}\\ \cline{2-6}
& 10Mb & 9.995 & 36.51 & 10.022 & 36.04 &\\ \cline{2-6}
& 20Mb & 20.133 & 39.47 & 20.024 & 38.90 &\\ \cline{2-6}
& 40Mb & 39.792 & 42.89 & 40.023 & 42.40 &\\
\hline
\multirow{4}{*}{I09} & 5Mb & 5.216 & 33.96 & 5.023 & 33.27 &\multirow{4}{*}{-17.14\%}\\ \cline{2-6}
& 10Mb & 10.031 & 36.18 & 10.022 & 35.53 &\\ \cline{2-6}
& 20Mb & 19.986 & 38.74 & 20.024 & 38.01 &\\ \cline{2-6}
& 40Mb & 39.952 & 41.94 & 40.024 & 40.98 &\\
\hline
\end{tabular}
}
\end{table}

 \begin{table}[!t]
\renewcommand{\arraystretch}{1.3}
\caption{Experimental Results on Rate-distortion Performances, Random-access, $\lambda$ = 2 (Bits in Mbits)}
\label{table_rdperform2}
\centering
\resizebox{0.48\textwidth}{!}{
\begin{tabular}{|c|c|c|c|c|c|c|}
\hline
\multirow{2}{*}{LFI} & \multirow{2}{*}{Bit budget} & \multicolumn{2}{c|}{proposed} & \multicolumn{2}{c|}{HM-RC} & \multirow{2}{*}{BD-rate}\\
\cline{3-6}
& & Bits & $T'$ & Bits & $T'$ & \\
\hline
\multirow{4}{*}{I01} & 1Mb & 0.987 & 37.29 & 0.996 & 36.49 & \multirow{4}{*}{-23.95\%}\\ \cline{2-6}
& 2Mb & 2.001 & 39.84 & 1.995 & 38.55 &\\ \cline{2-6}
& 4Mb & 3.999 & 41.93 & 4.021 & 41.18 &\\ \cline{2-6}
& 8Mb & 7.948 & 43.99 & 7.952 & 43.70 &\\
\hline
\multirow{4}{*}{I03} & 1Mb & 0.976 & 42.80 & 0.995 & 42.17 & \multirow{4}{*}{-14.48\%}\\ \cline{2-6}
& 2Mb & 1.979 & 44.22 & 1.996 & 43.87 &\\ \cline{2-6}
& 4Mb & 4.057 & 45.68 & 3.985 & 45.42 &\\ \cline{2-6}
& 8Mb & 7.846 & 46.95 & 7.855 & 46.68 &\\
\hline
\multirow{4}{*}{I07} & 1Mb & 0.990 & 35.00 & 0.972 & 32.83 & \multirow{4}{*}{-25.19\%}\\ \cline{2-6}
& 2Mb & 2.019 & 37.55 & 2.004 & 36.24 &\\ \cline{2-6}
& 4Mb & 4.061 & 39.73 & 4.069 & 38.93 &\\ \cline{2-6}
& 8Mb & 8.002 & 41.75 & 8.003 & 41.39 &\\
\hline
\multirow{4}{*}{I09} & 1Mb & 0.997 & 34.94 & 1.074 & 33.57 & \multirow{4}{*}{-24.89\%}\\ \cline{2-6}
& 2Mb & 2.013 & 37.24 & 2.137 & 36.50 &\\ \cline{2-6}
& 4Mb & 3.955 & 39.34 & 4.191 & 38.50 &\\ \cline{2-6}
& 8Mb & 8.039 & 41.55 & 8.085 & 41.25 &\\
\hline
\end{tabular}
}
\end{table}

 \begin{table}[!t]
\renewcommand{\arraystretch}{1.3}
\caption{Experimental Results on Rate-distortion Performances, Low-delay, $\lambda$ = 4 (Bits in Mbits)}
\label{table_rdperform3}
\centering
\resizebox{0.48\textwidth}{!}{
\begin{tabular}{|c|c|c|c|c|c|c|}
\hline
\multirow{2}{*}{LFI} & \multirow{2}{*}{Bit budget} & \multicolumn{2}{c|}{proposed} & \multicolumn{2}{c|}{HM-RC} & \multirow{2}{*}{BD-rate}\\
\cline{3-6}
& & Bits & $T'$ & Bits & $T'$ & \\
\hline
\multirow{4}{*}{I01} & 500Kb & 0.501 & 38.94 & 0.506 & 38.41 & \multirow{4}{*}{-21.05\%}\\ \cline{2-6}
& 1Mb & 0.995 & 40.09 & 1.006 & 39.60 &\\ \cline{2-6}
& 2Mb & 1.957 & 41.04 & 2.006 & 40.87 &\\ \cline{2-6}
& 4Mb & 3.876 & 42.09 & 4.007 & 41.92 &\\
\hline
\multirow{4}{*}{I03} & 500Kb & 0.506 & 43.64 & 0.506 & 42.70 & \multirow{4}{*}{-47.19\%}\\ \cline{2-6}
& 1Mb & 0.956 & 44.31 & 1.006 & 43.49 &\\ \cline{2-6}
& 2Mb & 1.810 & 44.90 & 2.006 & 44.69 &\\ \cline{2-6}
& 4Mb & 3.891 & 45.64 & 4.007 & 45.36 &\\
\hline
\multirow{4}{*}{I07} & 500Kb & 0.499 & 36.31 & 0.506 & 35.96 & \multirow{4}{*}{-17.18\%}\\ \cline{2-6}
& 1Mb & 0.998 & 37.55 & 1.006 & 37.31 &\\ \cline{2-6}
& 2Mb & 1.938 & 38.65 & 2.005 & 38.59 &\\ \cline{2-6}
& 4Mb & 3.885 & 39.87 & 4.006 & 39.90 &\\
\hline
\multirow{4}{*}{I09} & 500Kb & 0.505 & 36.53 & 0.506 & 36.21 & \multirow{4}{*}{-21.03\%}\\ \cline{2-6}
& 1Mb & 0.983 & 37.93 & 1.006 & 37.43 &\\ \cline{2-6}
& 2Mb & 1.952 & 39.06 & 2.006 & 38.78 &\\ \cline{2-6}
& 4Mb & 3.967 & 40.21 & 4.006 & 39.88 &\\
\hline
\end{tabular}
}
\end{table}
 
\begin{figure*}[!t]
\centering
\captionsetup{justification=centering}
\includegraphics[width=\textwidth]{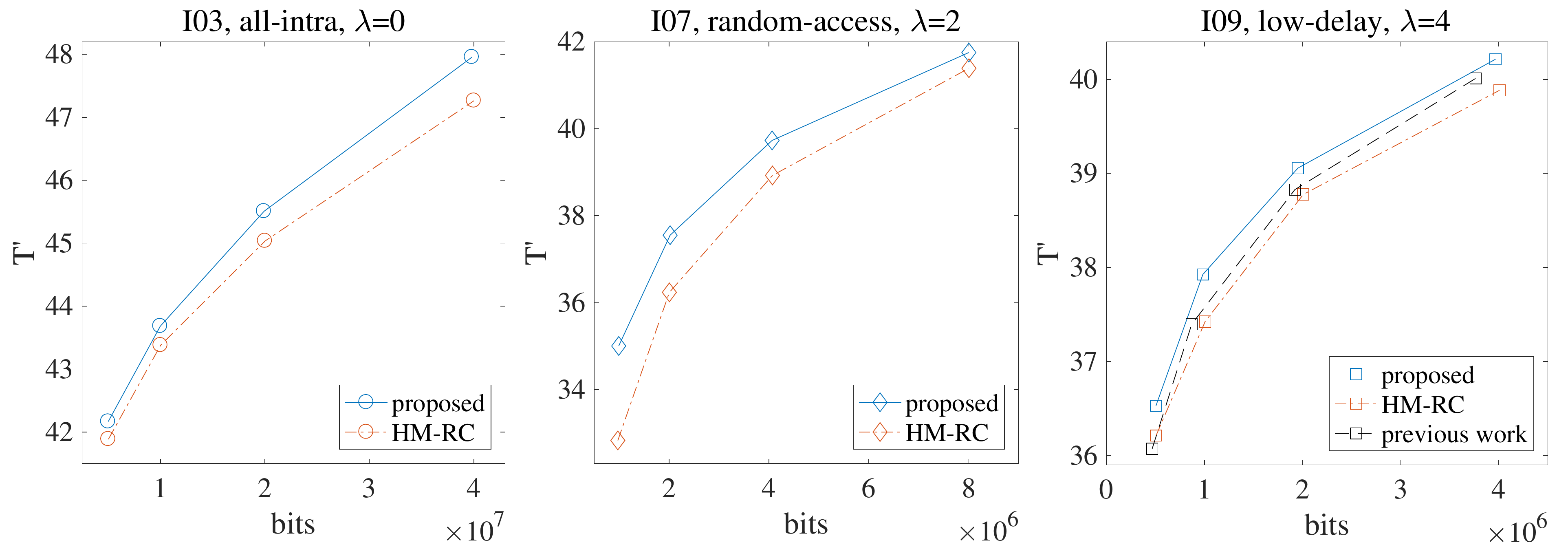}%
\caption{Example R-D curves of the algorithms.}
\label{fig:rdcurves}
\end{figure*}

\subsubsection{Rate-distortion performance}
\begin{figure*}[!t]
\centering
\captionsetup{justification=centering}
\includegraphics[width=\textwidth]{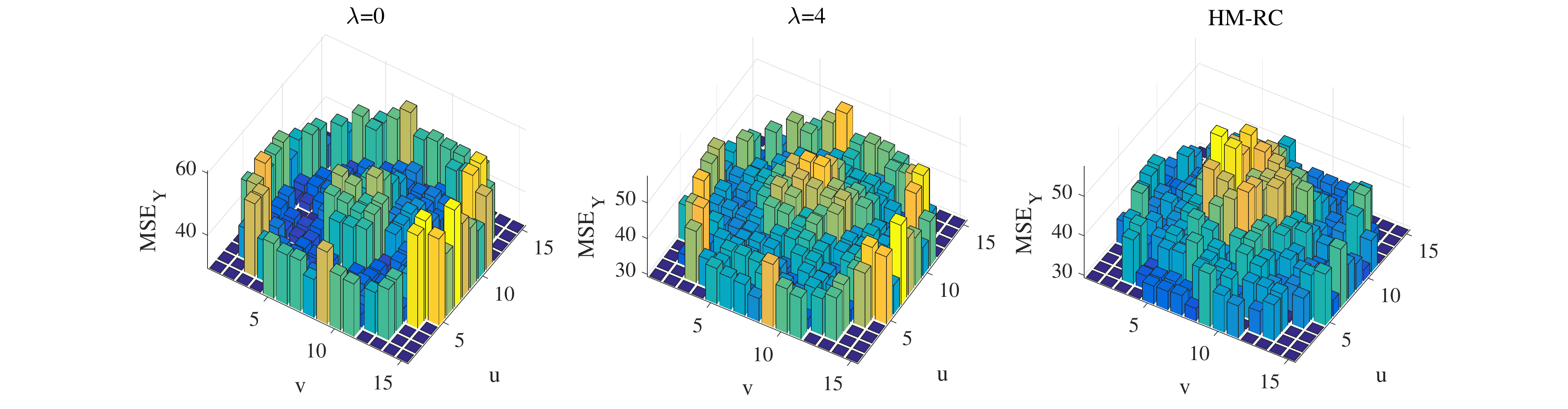}%
\caption{Illustration of the coding distortion distribution on the $uv$ plane. I09 is encoded with the low-delay configuration, where the target bit budget is set to 500Kb. Three different settings are tested: $\lambda$=0, $\lambda$=4, and the anchor.}
\label{fig:mse-ld}
\end{figure*}

 We use four LFIs as examples (I01, I03, I07, I09, all from different categories) to show the rate-distortion performance of
 the proposed two-pass encoding system.
 For comparison, the default rate-control algorithm in HM 16.16 was used as the anchor.
 Since the optimization target of the proposed system is $T$ defined by (\ref{eq:T-metric}),
 the rate-distortion performance is evaluated in terms of BD-rate \cite{Bjontegaard},
 where the PSNR is replaced by the following logarithmic form:
\begin{align}
 	T' = 10\log_{10}\bigg(\frac{255^2}{T}\bigg).
\end{align}
As explained in previous sections, the parameter $\lambda$ in (\ref{eq:T-metric})
controls the trade-off between spatial quality and angular consistency.
Therefore, three different levels of $\lambda$ were tested.
Table \ref{table_rdperform} shows the results encoded using the all-intra configuration,
 where $\lambda$ was set to 0 to exclude the angular consistency factor.
Table \ref{table_rdperform3} shows the results encoded using the low-delay configuration,
where $\lambda$ was set to 4 to yield results with smooth transitions among adjacent SAIs (see Fig.~\ref{fig:mse-ld}).
 Table \ref{table_rdperform2} shows the results encoded using the random-access configuration,
where $\lambda$ was set to 2 as an intermediate case.
The overall results, where all 13 LFIs were tested, are summarized in Table \ref{table_bdrates}.
All 3 encoder configurations were tested with 3 different levels of $\lambda$,
giving a total of 9 test settings.
 
 From Table \ref{table_rdperform}, \ref{table_rdperform2} and \ref{table_rdperform3}, we can find that
 the proposed two-pass encoding system is able to achieve higher $T'$ (or equivalently, reduce the optimization target) comparing to the default 
 rate-control algorithm in HM 16.16,
 without significantly increasing the bit costs.
 This improvement is consistent in all bit budgets and encoder configurations.
 Table \ref{table_bdrates} shows the average BD-rate improvement of all test settings.
 The proposed two-pass encoding system achieves 11.2\% to 11.9\% BD-rate reductions for the all-intra configuration, 
 15.8\% to 32.7\% BD-rate reductions for the random-access configuration,
 and 12.1\% to 15.7\% BD-rate reductions for the low-delay configuration.
 It can be seen that for the random-access and the low-delay configurations,
 the BD-rate improvement is higher for larger $\lambda$, where angular consistency is prioritized.
 This is due to the fact that the default rate-control algorithm in HM 16.16
 produces results with sharp transitions among adjacent SAIs, 
 and the proposed algorithm is able to mitigate this problem for better angular consistency.
 The average BD-rate improvement is higher for the random-access configuration comparing to the low-delay configuration,
 which can be explained by the difference of rate-distortion models. 
 The rate-distortion model used for the low-delay configuration is an approximation based on the virtual GOP assumption, 
 while that of the random-access configuration is based on actual GOPs.
Unlike the other two cases, the value of $\lambda$ does not significantly 
 affect the BD-rate improvement for the all-intra configuration.
 This is due to the fact that the default rate-control algorithm in HM 16.16
 for the all-intra configuration already produces homogeneous outputs,
 therefore the proposed algorithm offers little improvement on angular consistency.
 Also, the BD-rate improvement achieved on the all-intra configuration is relatively lower than the other two cases,
 which can be explained by that the homogenous outputs produced by the default 
 rate-control algorithm fit well to the plateau of the pixel confidence distribution in the center of $uv$ plane (see Fig.~\ref{fig:confidence}).
 
 In our previous work \cite{guo2018convex},
 we proposed an iterative encoding system for the low-delay configuration which
 ignores inter-frame dependency.
 The last three columns in Table \ref{table_bdrates} shows its performance 
 on the dataset,
 where the number of iterations was set to 3 as in \cite{guo2018convex}.
 Comparing to the iterative encoding system,
 the proposed two-pass encoding system improves the BD-rate reductions
 by 8.3\% ($\lambda$=0), 5.0\% ($\lambda$=2) and 3.1\% ($\lambda$=4).
 This mainly results from the improvement to the rate-distortion model,
 as in our previous work the rate-distortion behavior of each frame is
 assumed to be independent,
 while the virtual GOP based rate-distortion model (\ref{eq:rd-ra2}) is better
 at capturing the inter-frame dependency in the low-delay configuration.
 Some example R-D curves are shown in Fig.~\ref{fig:rdcurves}.
\begin{table}[!t]
\renewcommand{\arraystretch}{1.3}
\caption{Experimental Results on Rate-control Accuracy, Average Bit Errors}
\label{table_rcaccuracy}
\centering
\begin{tabular}{|c|c|c|c|}
\hline
Configuration & $\lambda$=0 & $\lambda$=2 & $\lambda$=4\\
\hline
all-intra  & 0.51\% & 0.75\% & 0.85\% \\ \hline
random-access  & 0.76\% & 0.88\% & 0.92\% \\ \hline
low-delay & 2.95\% & 1.70\% & 1.69\% \\ 
low-delay, algorithm \cite{guo2018convex} & 7.85\% & 7.39\% & 7.63\% \\ 
(improvement) & (-62.42\%) & (-77.00\%) & (-77.85\%) \\\hline
\end{tabular}
\end{table}

\subsubsection{Rate-control accuracy}
Although the main goal of the proposed two-pass encoding system
is to achieve a good balance between spatial quality and angular consistency,
it is also desirable that the size of the encoded result is close to the given bit budget.
The bit error can arise from the finite number of choices of frame QPs,
which causes the absolute value terms in (\ref{eq:abs-diffr1})(\ref{eq:abs-diffr2})
 to be non-zero.
 Statistical approximations, including the virtual GOP assumption for the low-delay configuration, the hyperbolic regression model,
 as well as the first order approximation (\ref{eq:rd-intra-taylor}),
 also contribute to the bit error.
Table \ref{table_rcaccuracy} shows the average bit errors of all 13 LFIs encoded
using the proposed two-pass encoding system as well as the iterative encoding system from our previous work,
categorized according to the encoder configurations and the value of $\lambda$ being used.
For the low-delay configuration, we see that the proposed two-pass encoding system reduces the average bit errors by 62.4\% to 77.9\% comparing to the previous work.
Again this improvement mainly originates from the better performance of virtual 
GOP based rate-distortion model (\ref{eq:rd-ra2}).
The achieved bit errors are lower for the random-access configuration, 
since the virtual GOP approximation is replaced by actual GOPs.
The proposed algorithm is able to achieve even lower bit errors for the all-intra configuration,
as the corresponding optimization (\ref{eq:opt-2}) operates on the frame level instead of the GOP level as in (\ref{eq:opt-3}), 
thereby offering better granularity at rate-control.
Although the proposed bit allocation framework has less granularity in terms of rate-control comparing to basic unit level rate-control algorithms such as \cite{li2014lambda},
the optimizations constrained by an upper bound  can still limit the bit errors below 0.85\% (frame level) or 2.95\% (GOP level),
depending on the level the optimizations are carried out on.

\begin{table}[!t]
\renewcommand{\arraystretch}{1.3}
\caption{Experimental Results on Encoding Time, Average Encoding Time Relative to the 1-pass Anchor}
\label{table_enctime}
\centering
\begin{tabular}{|c|c|c|c|}
\hline
Configuration & $\lambda$=0 & $\lambda$=2 & $\lambda$=4\\
\hline
all-intra  & 127.8\% & 127.9\% & 127.9\% \\ \hline
random-access  & 119.7\% & 119.5\% & 120.0\% \\ \hline
low-delay & 145.4\% & 145.5\% & 145.7\% \\ 
low-delay, algorithm \cite{guo2018convex} & 1799.1\% & 1789.9\% & 1788.3\% \\
(speedup) & (12.37x) & (12.30x) & (12.27x) \\ \hline
\end{tabular}
\end{table}

\begin{figure*}[!t]
\centering
\captionsetup{justification=centering}
\includegraphics[width=\textwidth]{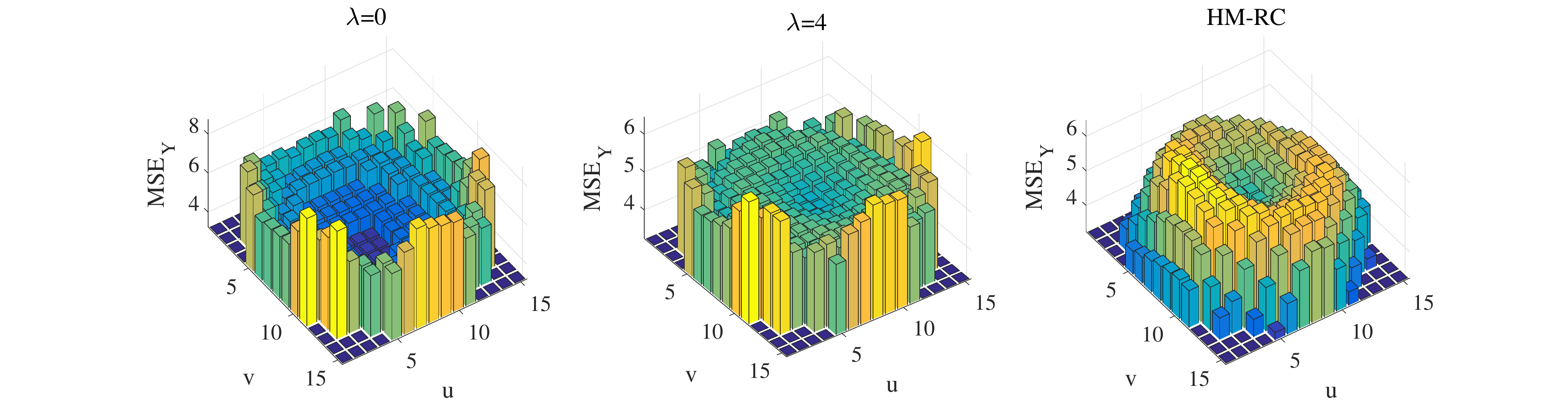}%
\caption{Illustration of the coding distortion distribution on the $uv$ plane. I03 is encoded with the all-intra configuration, where the target bit budget is set to 20Mb. Three different settings are tested: $\lambda$=0, $\lambda$=4, and the anchor.}
\label{fig:mse-intra}
\end{figure*}
\begin{figure*}[!t]
\centering
\captionsetup{justification=centering}
\includegraphics[width=\textwidth]{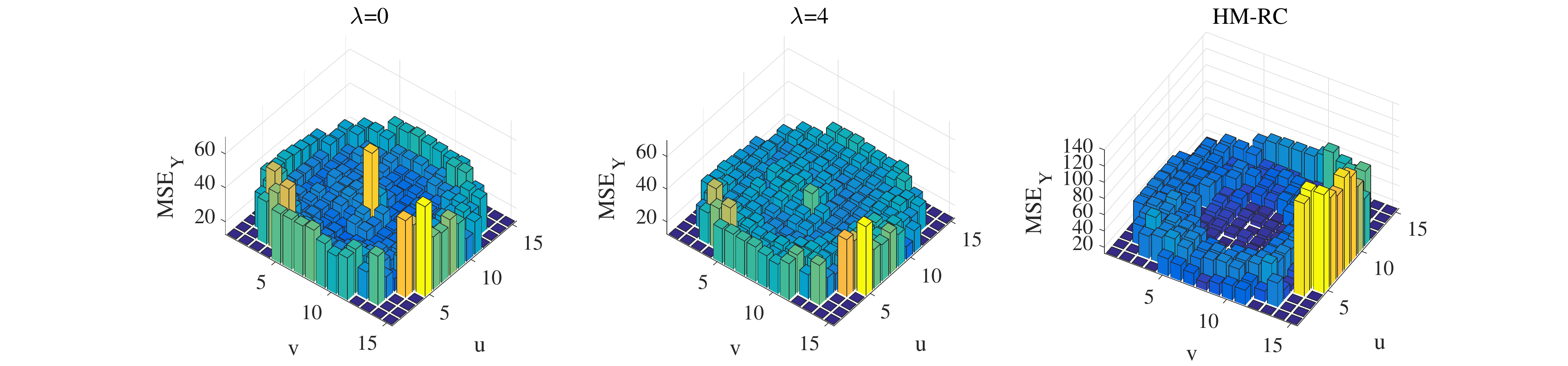}%
\caption{Illustration of the coding distortion distribution on the $uv$ plane. I07 is encoded with the random-access configuration, where the target bit budget is set to 2Mb. Three different settings are tested: $\lambda$=0, $\lambda$=4, and the anchor.}
\label{fig:mse-ra}
\end{figure*}

\begin{figure}[!t]
\centering
\includegraphics[width=0.4\textwidth]{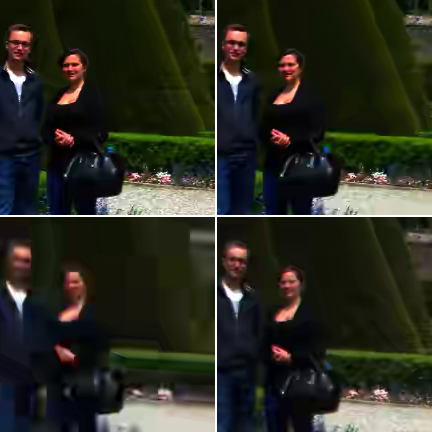}
\caption{Illustration of the effect of spatial quality adjustments. I09 is encoded with the all-intra configuration and a target bit budget of 10Mb. Top left: frame 4, $\lambda$=0. Bottom left: frame 179, $\lambda$=0. Top right: frame 4, anchor. Bottom right: frame 179, anchor.}
\label{fig:subj-sq}
\end{figure}
\begin{figure}[!t]
\centering
\includegraphics[width=0.4\textwidth]{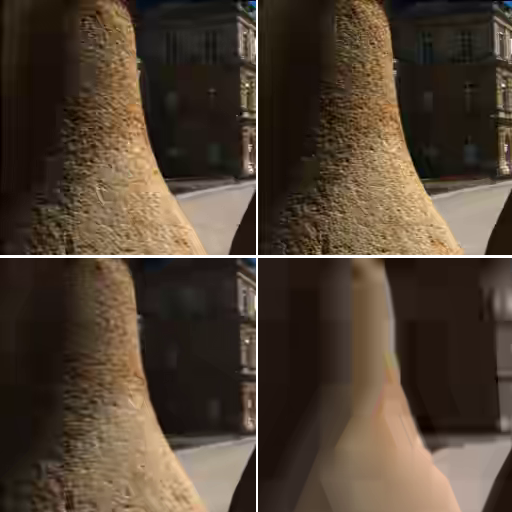}
\caption{Illustration of the effect of angular consistency adjustments. I13 is encoded with the random-access configuration and a target bit budget of 1Mb. Top left: frame 97, $\lambda$=4. Bottom left: frame 146, $\lambda$=4. Top right: frame 97, anchor. Bottom right: frame 146, anchor.}
\label{fig:subj-ac}
\end{figure}

\subsubsection{Encoding time}
A significant improvement of the proposed two-pass encoding system comparing to
the previous work is the encoding time reduction,
as the iterative serial trial compressions are replaced by a deterministic parallel two-pass encoding scheme.
Table \ref{table_enctime} shows the average encoding time of all 13 LFIs, 
relative to the encoding time using the 1-pass HM 16.16 anchor.
The time consumption of the proposed 2-pass encoding system includes the first pass encoding, convex optimization and the second pass encoding.
Note that for different target bit budgets, the first pass encoding is shared therefore only needs to be run once.
The algorithm from the previous work was also tested, where the
number of iterations was set to 3 as in \cite{guo2018convex}.
From Table \ref{table_enctime} we see that for the low-delay configuration,
the proposed two-pass encoding system is able to reduce the encoding time by more than
12x comparing to the previous work.
The iterative encoding scheme from \cite{guo2018convex} is very time-consuming,
as multiple passes of iterations are required, each iteration involves a serial trial
of multiple frame QPs.
In contrast, the time cost of the parallel first pass encoding in the proposed scheme  
is usually decided by the trial encoding with the lowest QP,
and the second pass is a simple fixed QP encoding.
Since the results of the first pass can be re-used for all target bit budgets,
the total time cost is even smaller. 
Table \ref{table_enctime} shows that for all test settings,
the relative time costs of the proposed two-pass encoding system 
are all less than twice of the one-pass anchor's time cost.
The exact values vary due to the encoder time cost with the lowest QP during the first pass and the target bit budgets are not equivalent for different encoder configurations.

\subsubsection{Coding distortion distribution}
To visualize the effects of using different levels of $\lambda$,
Fig.~\ref{fig:mse-intra} shows an example of the coding distortion distribution on the $uv$ plane.
The LFI I03 is encoded with the all-intra configuration under different settings ($\lambda$=0, $\lambda$=4, and the default rate-control in HM 16.16). The Y channel MSE is plotted against the $uv$ coordinates for each SAI.
It can be seen from the figure that when $\lambda$ is set to 0,
the optimization target ignores angular consistency and,
comparing to the anchor, 
the coding distortion is driven to external SAIs by the weighted MSE,
as those SAIs have relatively lower confidence.
When $\lambda$ is set to a positive value (e.g. 4), 
the smoothness penalty term induces the coding distortion to be more smoothly distributed than the case $\lambda$=0,
thereby promoting angular consistency.
This pattern demonstrates the effectiveness of the proposed optimization target.
Similar results are observed for other encoder configurations,
shown in Fig.~\ref{fig:mse-ra} and Fig.~\ref{fig:mse-ld}.

\subsubsection{Subjective quality}
The effectiveness of the proposed two-pass encoding system
can be visualized as follows.
For spatial quality, I09 is encoded using the proposed two-pass system with $\lambda$=0, and the anchor HM 16.16.
The encoded results are shown in Fig.~\ref{fig:subj-sq}.
Frame 4 is at the center of the $uv$ plane and frame 179 is at the edge of the $uv$ plane.
Since frame 179 has relatively lower confidence, strong optical distortion can be observed in the original SAI.
The figure shows that for frames with high confidence (e.g. frame 4), the proposed 
algorithm achieves fewer visual artifacts on human faces and the road.
The resulting increased bit cost is compensated by a reduced quality on frames with lower confidence (e.g. frame 179).
For angular consistency, I13 is encoded using the proposed two-pass system with $\lambda$=4, and the anchor HM 16.16.
The encoded results are shown in Fig.~\ref{fig:subj-ac}.
Frame 97 and frame 146 are two adjacent SAIs on the $uv$ plane.
It can be observed that the anchor yields results with strong discontinuity in visual quality,
while the proposed two-pass encoding system is able to yield results with much
smoother visual quality transitions.

\subsection{Summarization} \label{subsec:exp-sum}
From the results and discussions above, we can draw the following conclusions:
\begin{enumerate}
	\item The proposed two-pass encoding system can effectively improve the rate-distortion performance of LFI encoding to achieve better spatial quality and angular consistency for various encoder configurations;
	\item The proposed scheme for the low-delay configuration further improves the rate-distortion performance comparing to the previous work;
	\item The proposed scheme for the low-delay configuration significantly reduces the bit errors comparing to the previous work, and even lower bit errors are achieved on other encoder configurations;
	\item The proposed scheme for the low-delay configuration speeds up the previous work by more than 12x. The overall two-pass encoding system takes less time than twice of the one-pass anchor for all encoder configurations.
\end{enumerate}

\section{Conclusion} \label{sec:conclusion}
In this paper, we propose a novel frame level bit allocation framework for PTS-based
LFI compression guided by an optimization target that combines spatial quality and 
angular consistency.
The framework is carried out by a two-pass encoding system, 
where rate-distortion models are estimated during the first pass,
and the second pass performs the actual encoding with optimized bit allocations 
given by convex programming.
Experimental results show that comparing to the default rate-control of HM 16.16, the proposed two-pass encoding system on average achieves 11.2\% to 11.9\% BD-rate reductions for the all-intra configuration, 
15.8\% to 32.7\% BD-rate reductions for the random-access configuration,
and 12.1\% to 15.7\% BD-rate reductions for the low-delay configuration.
The resulting bit errors are limited (below 0.85\% for frame level optimizations and below 2.95\% for GOP level optimizations),
and the total time cost is less than twice of the one-pass anchor.
Comparing to the previous work \cite{guo2018convex} for the low-delay configuration,
the proposed two-pass encoding system achieves 3.1\% to 8.3\% further BD-rate improvement,
reduces the bit errors by more than 60\%, and speeds up the encoding processes by more than 12x.
Since this paper only discusses bit allocation optimizations on the frame level,
this leaves basic unit (CU) level optimizations for future work.


%

%
%
%
%
%
%
%

\ifCLASSOPTIONcaptionsoff
  \newpage
\fi

\bibliographystyle{IEEEtran}
\bibliography{./refs.bib}
%
%
%

%

\begin{IEEEbiography}[{\includegraphics[width=1in,height=1.25in,clip,keepaspectratio]{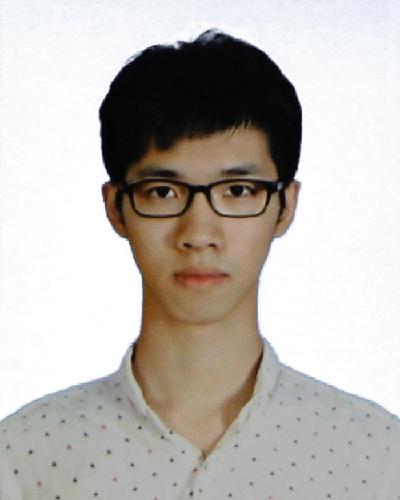}}]{Bichuan Guo}
received his B.S. degree from the Department of Electronic Engineering, Tsinghua University, Beijing, China in 2016, where he is currently in his 2nd year of the Ph.D. program with the 
Department of Computer Science and Technology. His research interests are video coding, panoramic video rendering and network data transmission optimization.
\end{IEEEbiography}

\begin{IEEEbiography}[{\includegraphics[width=1in,height=1.25in,clip,keepaspectratio]{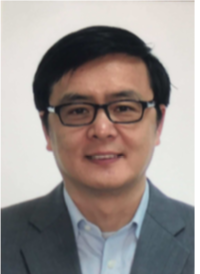}}]{Jiangtao Wen}
received the BS, MS and Ph.D. degrees with honors from Tsinghua University, Beijing, China, in 1992, 1994 and 1996 respectively, all in Electrical Engineering.

From 1996 to 1998, he was a Staff Research Fellow at UCLA, where he conducted research on multimedia coding and communications. Many of his inventions were later adopted by international standards such as H.263, MPEG and H.264. Some of his inventions were licensed to Samsung in the largest patent licensing agreement in the history of the highly ranked UCLA School of Engineering. After UCLA, he served as the Principal Scientist of PacketVideo Corp. (NASDAQ: WAVE/DCM), the CTO of Morphbius Technology Inc., the Director of Video Codec Technologies of Mobilygen Corp (NASDAQ: MXIM), the Senior Director of Technology of Ortiva Wireless (NASDAQ: ALLT) and consulted for Stretch Inc., Ocarina Networks (NASDAQ: DELL) and QuickFire Networks (NASDAQ: FB). Since 2009, Dr. Wen has held a Professorship at the Department of Computer Science and Technology of Tsinghua University. He was a Visiting Professor at Princeton University in 2010 and 2011.

Dr. Wen's research focuses on multimedia communication over challenging networks and computational photography. He has authored many widely referenced papers in related fields. Products deploying technologies that Dr. Wen developed are currently widely used worldwide. Dr. Wen holds over 40 patents with numerous others pending. Dr. Wen is an Associate Editor for IEEE Transactions Circuits and Systems for Video Technologies (CSVT). He is a recipient of the 2010 IEEE Trans. CSVT Best Paper Award and a nominee for the 2016 IEEE Trans CSVT Best Paper Award.

Dr. Wen was elected a Fellow of the IEEE in 2011. He is the Director of the Research Institute of the Internet of Things of Tsinghua University, and a Co-Director of the Ministry of Education Tsinghua-Microsoft Joint Lab of Multimedia and Networking.

Besides teaching and conducting research, Dr. Wen also invests in high technology companies as an angel investor.
\end{IEEEbiography}

\begin{IEEEbiography}[{\includegraphics[width=1in,height=1.25in,clip,keepaspectratio]{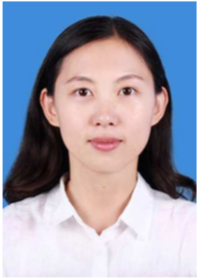}}]{Yuxing Han}
received her B.S from Hong Kong University of Science and Technology, and Ph.D from UCLA, both in Electrical Engineering. Yuxing is currently a professor in school of engineering at South China Agriculture University, China. Her research area focuses on virtual reality, multimedia communication over challenging networks and big data analysis. She has authored many widely referenced papers and patents in related fields. Products deploying technologies that Dr. Han developed are currently widely used worldwide. The high quality video encoding/transcoding solution project Yuxing led won 2016 Frost \& Sullivan best practice in Enabling Technology Leadership award. 
\end{IEEEbiography}

\end{document}